\documentclass[twocolumn,showpacs,amsfonts,aps,prd,floatfix,nofootinbib,float,superscriptaddress]{revtex4-1} 
\usepackage{graphicx}  
\usepackage{dcolumn}   
\usepackage{bm}        
\usepackage{amssymb}   
\usepackage{amsmath}   
\usepackage{tabularx}
 \usepackage[percent]{overpic}
\usepackage{placeins}
\usepackage{mathtools}
\usepackage{rotating}
\usepackage{color}

\newcommand{\rmii}[1]{{\mbox{\tiny\rm{#1}}}}
\newcommand{\beq}{\begin{equation}}
\newcommand{\eeq}{\end{equation}}
\newcommand{\bea}{\begin{align}}
\newcommand{\eea}{\end{align}}
\newcommand{\beas}{\begin{align*}}
\newcommand{\eeas}{\end{align*}}

\newcommand{\eps}{{\varepsilon}}
\newcommand{\mD}{m_\rmii{D}}

\newcommand{\vE}{\vec E}
\newcommand{\rh}{\hat r}
\renewcommand{\vr}{\vec r}
\newcommand{\vp}{\vec p}

\newcommand{\Tint}[1]{{\hbox{$\sum$}\!\!\!\!\!\!\!\int\,}_{\!\!\!\!\raise-0.9ex\hbox{$\scriptstyle{#1}$}}}
\renewcommand{\Re}{\rm Re}
\renewcommand{\Im}{\rm Im}

\definecolor{mygray}{gray}{0.6}

\begin{document}

\title{ Complex heavy-quark potential and Debye mass in a gluonic medium\\from lattice QCD}
\author{Yannis Burnier}
\affiliation{Institute of Theoretical Physics, EPFL, CH-1015 Lausanne, Switzerland}
\author{Alexander Rothkopf}
\affiliation{Institute for Theoretical Physics,  Heidelberg University, Philosophenweg 16, 69120 Heidelberg, Germany}

\date{\today}

\begin{abstract}
We improve and extend our study of the complex in-medium heavy quark potential and its Debye mass $m_D$ in a gluonic medium with a finer scan around the deconfinement transition and newly generated ensembles closer to the thermodynamic limit. On the lattices with larger physical volume, $\Re[V]$ shows signs of screening, i.e.\ a finite $m_D$, only in the deconfined phase, reminiscent of a genuine phase transition. Consistently $\Im[V]$ exhibits nonzero values also only above $T_C$. We compare the behavior of $\Re[V]$ with the color singlet free-energies that have been used historically to extract the Debye mass. An effective coupling constant is computed to assess the residual influence of the confining part of the potential at $T>0$. Our previous finding of a gradual screening of $\Re[V]$ around $T_C$ on finer lattices is critically reassessed and interpreted to originate from finite volume artifacts. We discuss that deficiency of the $\beta=7$, $\xi_b=3.5$ parameter set at $N_s=32$, which had been in deployed in the literature before. 
\end{abstract}

\maketitle

\flushbottom

\FloatBarrier

\section{Introduction}

At zero temperature the potential between a static quark-antiquark pair provides a powerful, yet intuitive handle on the physics of the strong interactions. From the inspection of this quantity we may already learn about the two central hallmarks of quantum-chromo-dynamics, i.e.\ asymptotic freedom and confinement. Evaluations of the potential using non-perturbative lattice QCD techniques have shown \cite{Koma:2006si} that it contains a Coulombic part at small distances, as well as a linearly rising part at larger distances corresponding to manifestations of these two phenomena respectively. In the presence of dynamical fermions, the linear rise will eventually terminate and go over to a constant, as the spontaneous creation of a light quark pair ruptures the color string originally connecting the static test color charges. Even details about the renormalization properties of QCD can be inferred as e.g.\ the running of the coupling observable at very small distances allows a clean connection with weak coupling computations \cite{Bazavov:2014soa,Pineda:2002se}. 

While it has been postulated for a long time \cite{Nadkarni:1986cz} that the physics of static quarks in a thermal medium might also be amenable to a simple description in terms of a non-relativistic in-medium potential, it took the maturation of effective field theory frameworks \cite{Brambilla:2004jw}, such as NRQCD and pNRQCD, to put this concept on a solid theoretical footing. Previously a popular approach was to model the potential \cite{Wong:2004zr,Kaczmarek:2005ui} by resorting to an identification with e.g.\ the color singlet free energies $F^{(1)}$
\begin{align}
\nonumber e^{\beta F^{(1)}(r)}=\Big\langle {\rm Tr}\Big[ \Omega(r)\Omega^\dagger(0) \Big]\Big\rangle_{\rm CG}\\ \Omega(r)={\rm exp}\Big[-ig\int_0^\beta d\tau A_\mu^a(r,\tau)T^a\Big],
\end{align}
defined from the correlator of two appropriately projected Polyakov loops $\Omega$. As different proposals for model potentials were put forward, among them also the color singlet internal energies, a long-lasting discussions ensued. These were only laid to rest after the groundbreaking work of Laine et. al. \cite{Laine:2007qy}, which showed through a computation in resummed hard-thermal loop perturbation theory that none of the purely-real model potentials can represent the actual in-medium potential as it must contain an imaginary part. The complex nature of the potential is intimately connected with the realization that it is an inherently Minkowski time quantity and cannot directly be computed from Euclidean observables accessible in e.g.\ lattice QCD. 

Here we provide a state-of-the-art determination of the in-medium potential in a gluonic medium using a recently established strategy to extract its values from a spectral decomposition of the thermal Wilson loop. In sec.\ref{sec:PotDef} we give a brief review of the definition of the in-medium potential and the strategy of how to extract it from lattice QCD simulations in Euclidean time. In addition we discuss the Gauss-law ansatz that has been developed to analytically capture the in-medium modification of the potential and relate it to a Debye mass $m_D$. Section \ref{sec:CoarseLat} describes in detail our new results on the in-medium potential in quenched QCD on coarse lattices closer to the thermodynamic limit, while sec.\ref{sec:FineLat} critically reassesses our previous results for the same quantity on finer lattices, however with smaller box sizes. The paper concludes with a summary and outlook in sec.\ref{sec:Conclusion}.

\section{Definition of the potential at $T>0$ and the Debye mass}
\label{sec:PotDef}

We briefly review here the definition of the in-medium static potential and its extraction from lattice QCD simulations, as well as our proposal for a gauge invariant definition of a Debye screening mass $m_D$.

The heavy quark potential arises in the language of effective field theory (EFT) as Wilson coefficient \cite{Brambilla:2004jw} of a non-relativistic framework for the description of heavy-quark bound states called pNRQCD \cite{Brambilla:1999xf,Brambilla:2008cx}. In this EFT the heavy quark two-body system is described in terms of color singlet and octet wavefunctions that evolve according to Schr\"odinger like equations. The matching of the static color singlet sector between pNRQCD and QCD amounts to identifying the correlator of the singlet wavefunctions in the latter with a correlation function of equal physical content in the latter, i.e.\ the thermal Wilson loop in the $m\to\infty$ limit

\begin{widetext}
\begin{eqnarray}
\nonumber \Big\langle \psi_S({\bf r},t)\psi_S({\bf r},0) \Big\rangle_{\rm pNRQCD} = D^>({\bf r},t)\overset{m\to\infty}{\equiv}  W_\square({\bf r},t)=\Big\langle {\rm Tr} \Big( {\rm exp}\Big[-ig\int_\square dx^\mu A_\mu^aT^a\Big] \Big) \Big\rangle_{\rm QCD}.
\label{Eq:ForwProp}
\end{eqnarray}
\end{widetext}

It is known that the Wilson loop in general obeys the following equation of motion \cite{Laine:2007qy}
\beq
i\partial_tW_\square(r,t)=\Phi(r,t)W_\square(r,t),
\eeq
where $\Phi(r,t)$ denotes a time- and space dependent complex function. If indeed a potential description of the two-body $Q\bar{Q}$ system is applicable the function $\Phi$ will asymptote towards a time independent quantity at late times, which we identify with the potential as
\beq
V(r)=\lim_{t\to\infty} \frac{i\partial_t W(t,r)}{W(t,r)}\label{Eq:VRealTimeDef}.
\eeq
This definition  of the static potential is formulated in real-time, which poses a difficulty to its evaluation using lattice QCD simulations that are conventionally performed in Euclidean or imaginary time. As has been proposed in \cite{Rothkopf:2009pk,Rothkopf:2011db} we may use a spectral decomposition to relate the Wilson loop in real-time with its Euclidean counterpart, which can be simulated numerically
\begin{align}
 \nonumber W_\square(\tau,r)=\int d\omega e^{-\omega \tau} \rho_\square(\omega,r)\,
&\leftrightarrow\, \\ \int d\omega e^{-i\omega t} \rho_\square(\omega,r)= W_\square(t,r).&
\end{align}
In turn the values of the potential itself can be related to the spectrum of Wilson loops
\begin{align}
V(r)=\lim_{t\to\infty}\int d\omega\, \omega e^{-i\omega t} \rho_\square(\omega,r)/\int d\omega\, e^{-i\omega t} \rho_\square(\omega,r). \label{Eq:PotSpec}
\end{align}

Extracting spectral information from lattice QCD however poses an ill-defined problem, as a continuous function needs to be estimated from a finite and noisy set of simulation data. One strategy to give meaning to this task is to deploy Bayesian inference that provides systematic means to incorporate additional prior information available on the spectrum. Early attempts to reconstruct the Wilson loop spectrum based on the popular Maximum Entropy Method \cite{Asakawa:2000tr,Rothkopf:2012vv} encountered difficulties due to e.g.\ the choice of search space in the state-of-the-art implementation by Bryan. It actually required the development of a novel Bayesian reconstruction prescription \cite{Burnier:2013nla} to achieve quantitatively robust reconstruction results.

Even if the spectrum is reliably reconstructed the necessity to take the late time limit remains. The need to actually carry out the Fourier transform can be circumvented by obtaining a better understanding of which spectral structures actually encode relevant information on the potential. Using the symmetries of the Wilson loop it was shown \cite{Burnier:2012az} that if a potential description exists its spectrum must contain a well defined lowest lying peak of skewed-Lorentzian type, whose position and width are directly related to the real- and imaginary part of the potential respectively
\begin{align}
&\rho\propto\frac{|{\rm Im} V(r)|{\rm cos}[{\rm Re}{\sigma_\infty}(r)]-({\rm Re}V(r)-\omega){\rm sin}[{\rm Re} {\sigma_\infty}(r)]}{ {\rm Im} V(r)^2+ ({\rm Re} V(r)-\omega)^2}\label{skewedrho}\\ \notag&+{c_0}(r)+{c_1}(r)({\rm Re} V(r)-\omega)+{c_2}(r)({\rm Re} V(r)-\omega)^2\ldots. 
\end{align}
I.e. after reconstructing the spectrum from lattice simulations we perform a fit of the lowest lying structure with the above functional form to extract the values of the in-medium potential. 

In practice we do not consider the Wilson loop but instead the correlator of Wilson lines, fixed to Coulomb gauge, as the latter does not possess the cusp divergencies that make the evaluation of the Wilson loop prohibitively expensive on fine lattices. In turn the spectrum of the Wilson line correlator is much more weakly populated compared to the Wilson loop at high frequencies, both improving the possibility of successful reconstruction of the spectral peak related to the potential and diminishing the influence of lattice spacing artifacts that affect the UV regime.

To quantitatively describe the in-medium modification of both $\Re[V]$ and $\Im[V]$ we propose to deploy an approach based on the generalized Gauss law derived in ref.~\cite{Dixit:1989vq}. It is formulated using the auxiliary vector field $\vE$ we may associate with the static singlet potential derived above. At zero temperature it reads
\beq
\vec\nabla \left(\frac{\vE}{r^{a+1}}\right)=   4\pi \,q\, \delta(\vr) \label{Eq:GenGauss}.
\eeq
To be more specific, in case of heavy quarkonium we assume that  the auxiliary (color) electric field $\vE=q r^{a-1} \rh$ either arises from the Coulombic $a=-1, q=\tilde\alpha_s, [\tilde\alpha_s]=1$ or the string-like part $a=1, q=\sigma, [\sigma]={\rm GeV}^2$ of the naive $T=0$ Cornell potential. We thus have three vacuum parameters that enter at this point, the strong coupling $\alpha_S$, the vacuum string tension $\sigma$ and a possible constant contribution $c$.

Our strategy to incorporate in-medium effects, detailed in \cite{Burnier:2015nsa}, relies on Fourier transforming the $T=0$ expression and multiplying the right hand side with the complex permittivity of a weakly coupled gas of light quarks and gluons 
\beq
\eps^{-1}(\vp,m_D)=\frac{p^2}{p^2+m_D^2}-i\pi T \frac{p\, m_D^2}{(p^2+m_D^2)^2}.\label{Eq:HTLperm}
\eeq
By this construction we attempt to separate the non-perturbative effects related to confinement which appear as linear contribution to the potential at $T=0$ from thermal effects that might be already well described by resummed hard-thermal loops. For the Coulombic part of the in-medium potential we recover the expressions of Laine et. al. \cite{Laine:2007qy} with a Yukawa type screening in the real part, as well as a finite imaginary part
\beq
V_c(r)= -\tilde\alpha_s\left[\mD+\frac{e^{-\mD r}}{r}
+iT\phi(\mD r)\right] \label{Eq:VHTL},
\eeq
where 
\beq
\phi(x)=2 \int_0^\infty dz \frac{z}{(z^2+1)^2}\left(1-\frac{\sin(xz)}{xz}\right).\label{phi}
\eeq
The in-medium modification of the string-like potential leads us to the expression
\begin{align}
{\rm Re}V_s(r)&=-\frac{\Gamma[\frac{1}{4}]}{2^{\frac{3}{4}}\sqrt{\pi}}\frac{\sigma}{\mu} D_{-\frac{1}{2}}\big(\sqrt{2}\mu r\big)+ \frac{\Gamma[\frac{1}{4}]}{2\Gamma[\frac{3}{4}]} \frac{\sigma}{\mu}
\end{align}
for the real part, where the relevant quantity that governs the strength of in-medium modification is $\mu^4=m_D^2\frac{\sigma}{\alpha_S}$. For the imaginary part we find 
\begin{eqnarray}
\Im V_s(r)&=&-i\frac{\sigma m_D^2 T}{\mu^4} \psi(\mu r)=-i\tilde\alpha_s T \psi(\mu r)\label{Eq:ImVSGenGauss},
\end{eqnarray}
where the term $\psi$ arises from a Wronskian construction
\begin{eqnarray} 
 \notag\psi(x)&=&D_{-1/2}(\sqrt{2}x)\int_0^x dy\, {\rm Re}D_{-1/2}(i\sqrt{2}y)y^2 \varphi(ym_D/\mu)\\\notag&&\hspace{-4mm}+{\rm Re}D_{-1/2}(i\sqrt{2}x)\int_x^\infty dy\, D_{-1/2}(\sqrt{2}y)y^2 \varphi(ym_D/\mu)\\ \nonumber&&-D_{-1/2}(0)\int_0^\infty dy\, D_{-1/2}(\sqrt{2}y)y^2 \varphi(ym_D/\mu).
\end{eqnarray}
In the end we simply form the sum of the Coulombic and string contribution to represent the full in-medium potential $V=V_c+V_s$\footnote{An approach in which the potential was directly modified with an in-medium permittivity has been proposed in \cite{Kakade:2015xua}, where however both a remnant long range component in $\Re[V]$ and a diverging imaginary part was observed}. The form of the solution of $\Re[V]$ bodes well for the application of this ansatz to temperatures around the phase transition, where HTL alone is not reliable anymore. I.e. we can take the limit $m_D\to0$ straight forwardly and recover the vacuum potential of Cornell type. Note that all in-medium modification of the potential is governed by a single temperature dependent parameter $m_D$. Apriori it is not clear whether this simple construction will allow us to reproduce the values of the potential obtained on the lattice. In the following sections we will however see examples where the Gauss-law ansatz succeeds very well in retracing $\Re[V]$. Another important aspect of our approach is that, once $m_D$ is fixed from a fit to $\Re[V]$ we can pre- or postdict the corresponding values of $\Im[V]$, which serves as crosscheck, since spectral widths are much more difficult to reconstruct with Bayesian methods than peak positions.

\section{New results on coarser lattices $\beta=6.1$ at $L=3.1$fm}
\label{sec:CoarseLat}

As in our previous study of quenched lattice QCD, we resort here to a fixed-scale approach.  I.e. temperature is changed between ensembles by varying the number of points in temporal direction, while keeping the lattice spacing fixed. In turn we only need to simulate a single low temperature reference from which e.g.\ the vacuum parameters for the Gauss-law fits are determined. We deploy the naive Wilson plaquette action with a combination of one standard multi-hit heat bath update together with five overrelaxation steps, allowing for a relatively long burn in period of 40000 such sweeps.  Subsequently a configuration is written out after 200 sweeps each. 

For the new ensembles closer to the thermodynamic limit we deploy a parameter set, for which the authors of ref.~\cite{Matsufuru:2001cp} have carried out a high precision scale setting. The anisotropic lattices with inverse coupling $\beta=6.1$ and bare anisotropy parameter $\xi_b=3.2108$ exhibit a physical lattice spacing of $a_s=0.097{\rm fm}=4a_\tau$. The naive estimates for the spatial cutoffs present on these lattices correspond to $\Lambda_{\rm UV}=11.1$GeV and $\Lambda_{\rm IR}=0.399$GeV. With our choice of $N_s=32$ points along the spatial axes a physical length of $L=3.1$fm ensues. To scan the temperature range around the deconfinement transition at $T_{C}^{\beta=6.1}=290$MeV we selected eight temporal axis extents $N_\tau=20-36$  corresponding to $T=1.4-0.78T_C$, as well as $N_\tau=72$ as a low temperature reference, compiled in Tab.~\ref{Tab:LatParmSU3-Coarse}.

\begin{table*}[t]
\small
\begin{tabularx}{15.5cm}{ | c | X | X | X | X | X | X | X | X | X |}
\hline
        $N_\tau$ & 20 & 22 & 24 & 26 & 28 & 30 & 32 & 36 & 72\\ \hline
	$T$[MeV] & 406 & 369 & 338 & 312 & 290 & 271 & 254 & 226 & 113 \\ \hline
	$T/T_C$ & 1.4 & 1.27 & 1.67 & 1.08 & $\approx1$ & 0.93 & 0.88 & 0.78 & 0.39 \\ \hline
	$N_{\rm meas}$ & 2080 & 3250 & 1920 & 1950 & 1730 & 900 & 880 & 940 & 950 \\ \hline
\end{tabularx}\caption{Quenched lattice ensembles at $\beta=6.1,\xi_b=3.2108, N_s=32$ corresponding to a physical $a=0.097{\rm fm}$ and $L=3.1{\rm fm}$. }\label{Tab:LatParmSU3-Coarse}
\end{table*}

\begin{figure}
\centering
\includegraphics[scale=0.4, trim=0cm 2cm 0.1cm 2cm, clip=true]{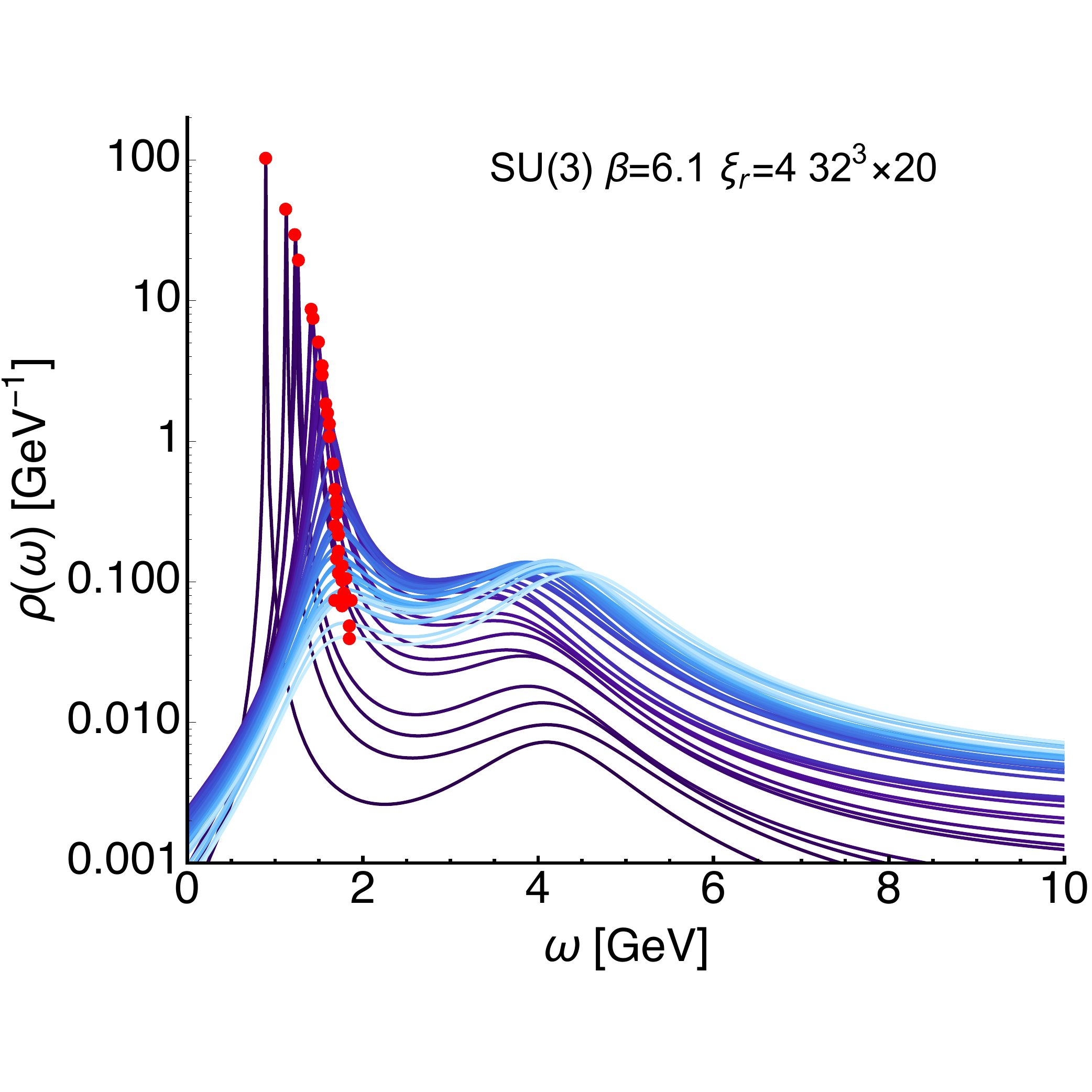}
\includegraphics[scale=0.4, trim=0cm 2cm 0.1cm 2cm, clip=true]{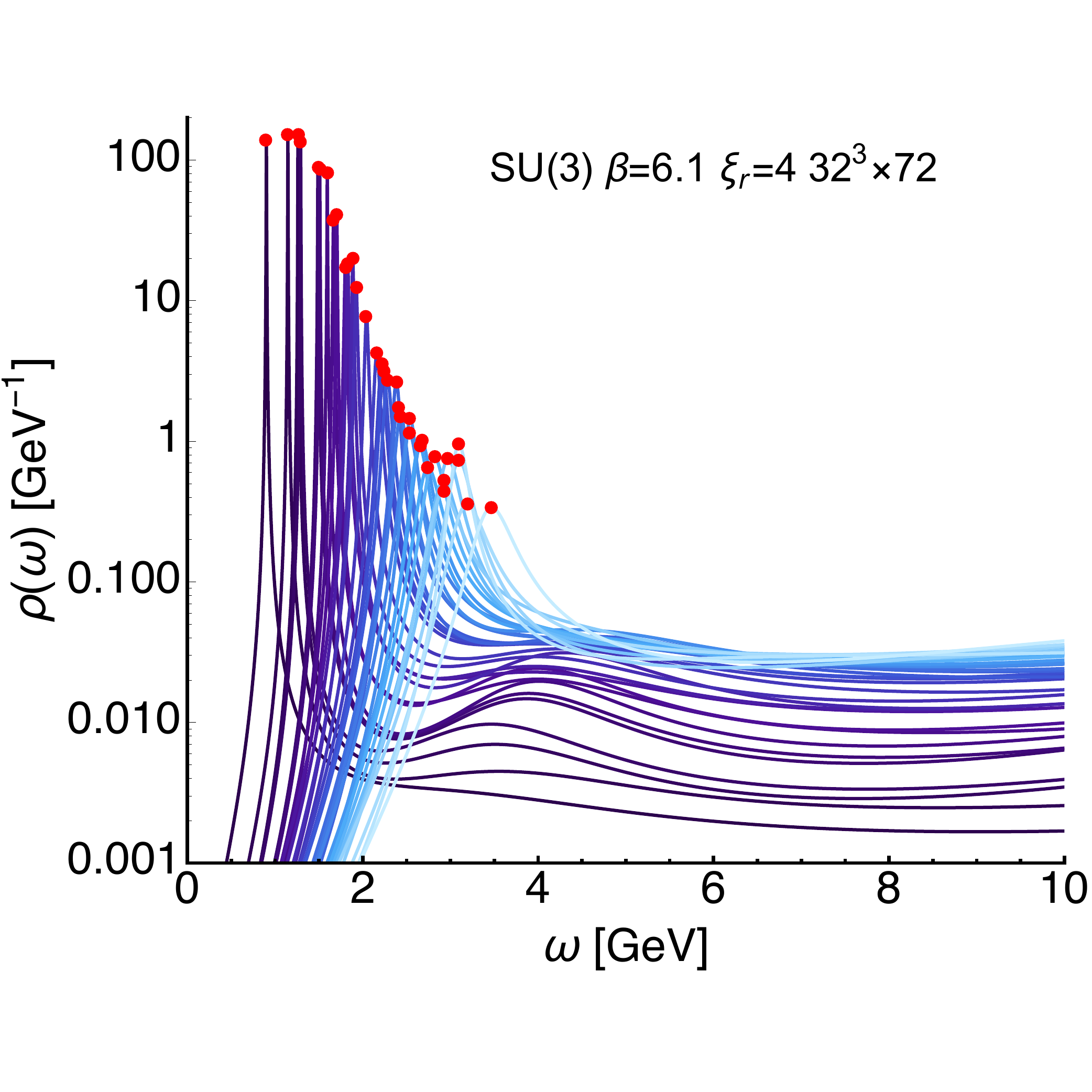}
\caption{Spectral reconstructions for the highest  $T=1.4T_C$ ($N_\tau=20$) (top) and the lowest temperature $T=0.39T_C$ ($N_\tau=76$) (bottom) at all available spatial separation distances $r/a\in[1,17]$. The lowest lying skewed-Lorentzian peak encoding the values of the in-medium potential is marked with a red circle. }\label{Fig:b61SpecRecs}
\end{figure}

The Wilson line correlators, which underlie our extraction of the potential are measured after fixing to Coulomb gauge with a tolerance of $\Delta_{\rm GF}=10^{-15}$. To lessen finite volume effects we measure their values both on-axis, as well as off-axis only along the $\sqrt{2}$ and $\sqrt{3}$ diagonals. The simulation data obtained is subjected to a Bayesian spectral reconstruction based on the recently developed BR method \cite{Burnier:2013nla}. For it we choose a common discretization of the numerical frequency interval of $\omega_{\rm num}\in[-5,5]$  with $N_\omega=3600$ points, while the $N_\tau$ input data are distributed along a rescaled $\tau$ interval $\tau_{\rm num}=[0,\beta_{\rm num}=20]$. We discard the first and last point in order to avoid contamination of the spectrum from overlap divergencies. After providing a constant default model with $m(\omega)=1$ we obtain a unique Bayesian reconstruction. Statistical errors of the reconstruction are estimated by a binned Jackknife procedure, where the reconstruction is carried out ten times, each time removing another interval of 10\% of the correlator measurements. The systematic uncertainties, in particular the dependence on the choice of default model are assessed by both varying its amplitude over two orders of magnitude, as well as its functional form as $m(\omega)=m_0(\omega-\omega_{\rm min}+1)^\xi$ with $\xi\in\{0,1,-1,-2\}$. In addition we repeat the reconstruction after removing 10\% of the small $\tau$ and/or large $\tau$ datapoints. The spectral reconstructions for all available values of $r/a\in[1,17]$ at high $T=1.4T_C$ ($N_\tau=20$, top) and low $T=0.39T_C$ ($N_\tau=72$, bottom) temperature are shown in Fig.~\ref{Fig:b61SpecRecs} \footnote{Formally a lattice spectral function is composed of a finite number of discrete delta peaks. Their spacing at sufficiently large physical volumes may however become fine enough to be regarded as continuous for the practical purpose of a spectral determination and we did not find indications of their discreteness impacting our reconstruction.}. Note that for all spatial distances a well defined lowest lying peak of skewed-Lorentzian type is found in the reconstructed spectra, its tip being marked by a red circle. 

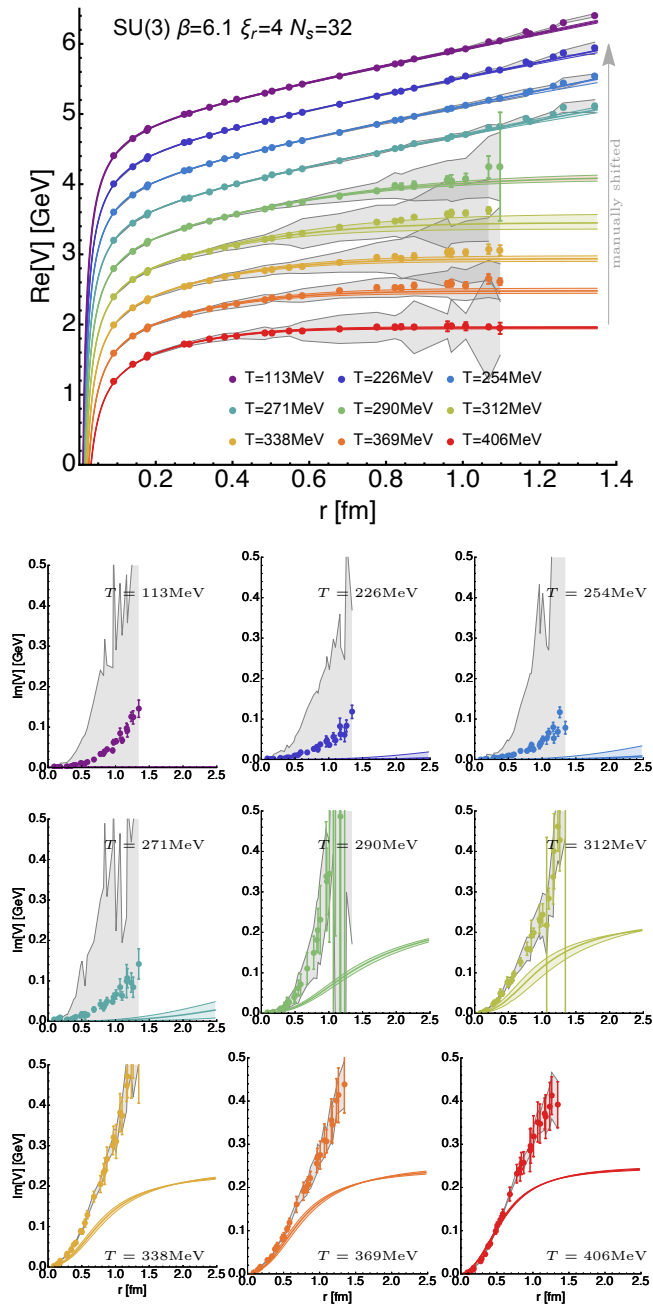
\begin{figure}
\centering
\begin{overpic}[scale=0.38,trim=0cm 1.cm 0cm 0.8cm, clip=true]{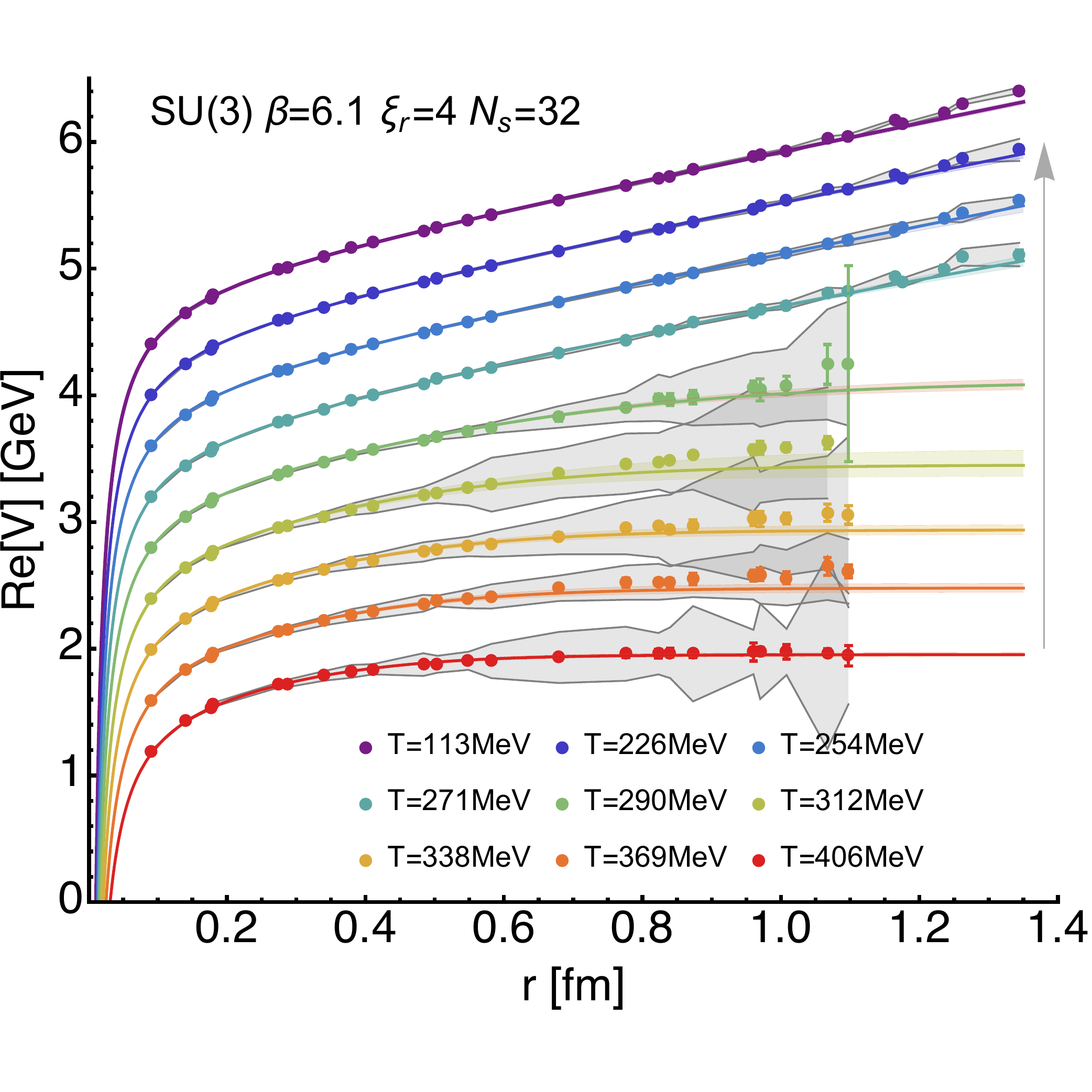}
\put (96,45) {\begin{turn}{90} \color{mygray}\tiny manually shifted \end{turn}}
\end{overpic}
\begin{overpic}[scale=0.19,trim=0cm 0cm 0cm 0cm, clip=true]{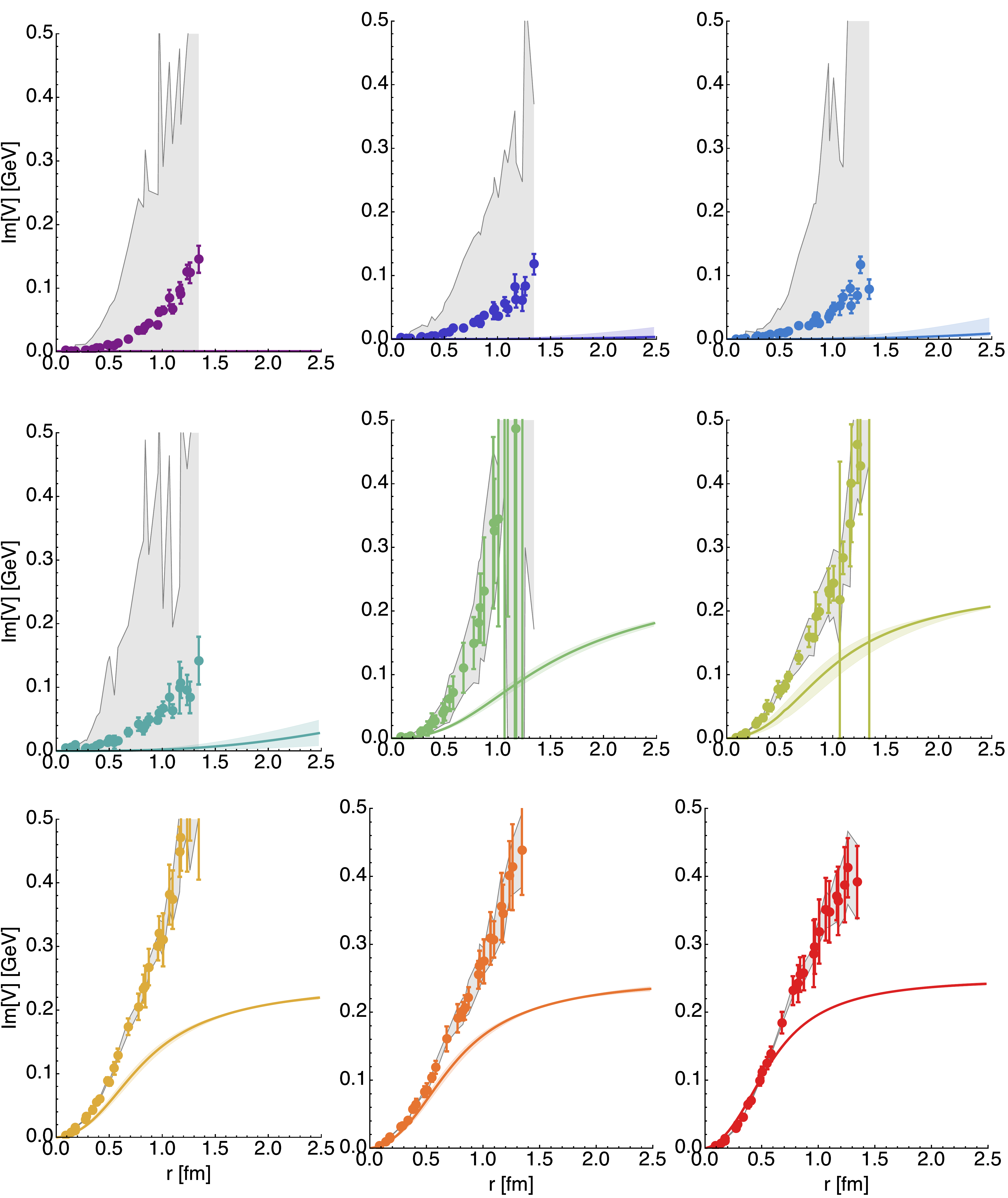}
 \put (70,6) {\tiny$T=406$MeV}  \put (40,6) {\tiny$T=369$MeV}  \put (12,6) {\tiny$T=338$MeV}
  \put (70,60) {\tiny$T=312$MeV} \put (40,60) {\tiny$T=290$MeV} \put (12,60) {\tiny$T=271$MeV}
 \put (70,93) {\tiny$T=254$MeV} \put (40,93) {\tiny$T=226$MeV} \put (12,93) {\tiny$T=113$MeV}
\end{overpic}
\caption{Extracted values of the real (top) and imaginary part (bottom) of the in-medium potential (colored points) at $\beta=6.1$. The values of $\Re[V]$ are shifted manually in y-direction for better readability as indicated by the gray arrow. The values of $\Im[V]$ are plotted individually as grid from lowest temperature (top left box) to highest temperature (bottom right box). The colored errorbars denote statistical uncertainty from a ten-bin Jackknife, while the gray errorband arises from varying both the default model and the $\tau$ interval of the underlying correlators. The Gauss-law fit to $\Re[V]$ on the top and the corresponding prediction of $\Im[V]$ on the bottom are plotted as solid colored lines. The colored errorbands arise from the uncertainty in the determination of the Debye mass parameter $m_D$.}\label{Fig:b61ReVImV}
\end{figure}

As laid out in sec.\ref{sec:PotDef} we fit the position and width of the lowest lying peak of each reconstructed spectrum in order to determine the values of the in-medium $\Re[V]$ and $\Im[V]$ at the corresponding spatial separation distance. Small values of $r$ on the lattice are known to suffer from finite lattice spacing artifacts, which we correct for using the prescription of ref.~\cite{Necco:2001xg} before plotting the values of the in-medium potential in Fig.~\ref{Fig:b61ReVImV} as colored points.

In the top panel of Fig.~\ref{Fig:b61ReVImV} we show the values of $\Re[V]$ manually shifted in y-direction for better readability as indicated by the gray arrow.  Statistical errors from the ten-bin Jackknife are given as colored errorbars. The reduction of the number of available points in $\tau$ direction when increasing the temperature in the fixed-scale approach leads to a growth of the systematic errors shown as gray bands. Already from an inspection by eye we find that the four top-most series of points below the deconfinement temperature $T_{C}^{\beta=6.1}=290$MeV show a clear linear behavior at large distances and only at higher temperatures a flattening off sets in. This behavior is reminiscent of a first order phase transition, as expected in SU(3) Yang-Mills in the thermodynamic limit, where the Polyakov loop shows a non-analytic jump to finite values as one crosses into the deconfined phase.

Consistent with this picture we also find that the values of the imaginary part we obtain are only significantly non-zero above $T_C$, as can be seen from the gray systematic errorbands. Even though the spectral reconstruction with one particular choice of default model gives a finite value of $\Im[V]$ at lower temperatures, varying $m(\omega)$ induces change in the related spectral width, which goes far beyond the statistical uncertainty there. At high temperature we expect that the reconstruction of $\Im[V]$ is reliable at small distances $r\sim0.3$fm and due to the relatively small number of available underlying correlator points however will quickly lead to an overestimation of the actual value.

To more quantitatively describe the in-medium modification we continue to investigate $\Re[V]$ with the help of the Gauss-law ansatz described in sec.\ref{sec:PotDef}. As a first step we need to determine the vacuum potential parameters, i.e.\ the strong coupling $\alpha_S$, the string tension $\sigma$ and the overall additive constant $c$. In absence of a true $T=0$ ensemble we use here the values from $N_\tau=72$ at $T=0.39T_C$, which in light of very similar behavior of $\Re[V]$ below $T_C$ appears justified. Fitting the corresponding Cornell potential form yields
\begin{align}
\nonumber &\alpha_S=0.272\pm0.016,\quad \sigma=0.215\pm0.003{\rm GeV}^2\\ &c=1.38\pm0.014{\rm GeV}
\end{align}
with the errors obtained from varying the fitting range by up to six steps at the upper and lower end of the fitting interval. The fit result is shown as the top violet curve on the top of Fig.~\ref{Fig:b61ReVImV}. Once the parameters are set, the Gauss-law ansatz contains only a single temperature dependent parameter, the Debye mass $m_D$, which controls the in-medium modification of both $\Re[V]$ and $\Im[V]$. Adjusting $m_D$ we are able to retrace the datapoints obtained from the lattice, as indicated by the colored solid lines in the top panel of Fig.~\ref{Fig:b61ReVImV}.  The agreement with $\Re[V]$ is excellent at all $r$, while the agreement with $\Im[V]$ is only satisfactory at high temperature and small distances. 

There are two factors that can contribute to the observed differences in Im[V]. On the one hand the extraction of the thermal width underlying the value of $\Im[V]$ is very challenging. Even though the position of spectral peaks may be well constrained based on correlators along $N_\tau\in[24..96]$ Euclidean points, previous studies suggest \cite{Burnier:2013nla} that the width may not yet have been captured satisfactorily. While for the Maximum Entropy Method the correct width is always approached from above the BR method used here may underestimate the width when applied to a small number of available datapoints evaluated at high precision. I.e. at small distances in r where the signal to noise ratio is high, the reconstructed width may even be smaller than the correct width, while with increasing r and concurrent lowering of the signal to noise ratio the width will become overestimated. Note that the dependence of the reconstruction on the number of available datapoints enters the systematic errorbands, in which cuts of up to 10\% of the correlator points from the low and large $\tau$ regime are considered.

\begin{figure}
\centering
\includegraphics[scale=0.39,trim=3cm 0cm 0cm 0cm, clip=true]{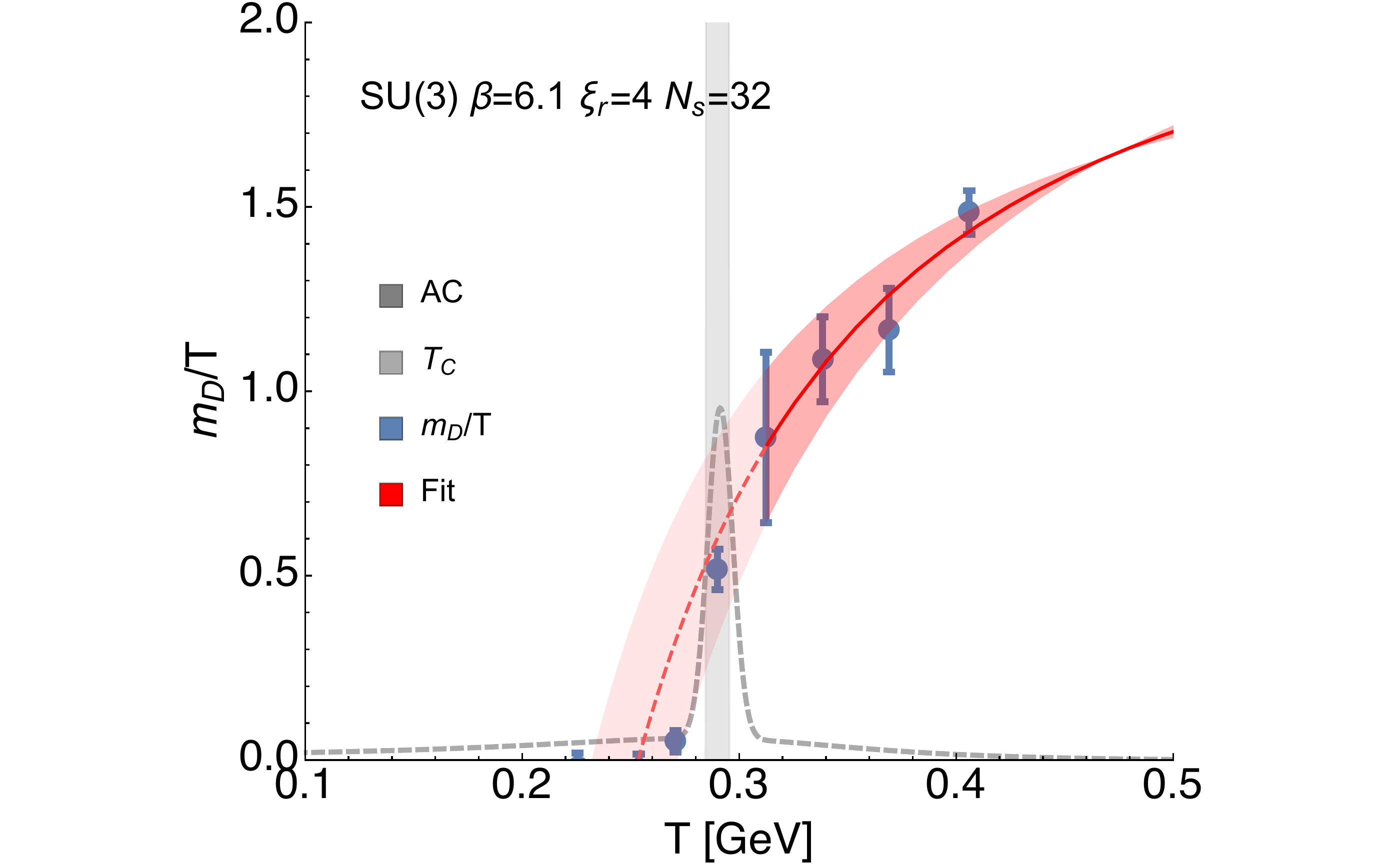}
\caption{Debye mass parameter of the heavy quark potential for $\beta=6.1$, describing its in-medium modification. The blue points denote the extracted values for $m_D/T$, while the red curve corresponds to a fit based on HTL perturbation theory amended by non-perturbative correction terms that allow for the non-perturbative downward trend around $T_C$. The transition temperature determined via the Polyakov loop is given as a gray band. Complementarily we show the autocorrelation time of the Wilson line correlators at $r/a_s=5.2$ and $\tau/a_\tau=14$ normalized to unity at its maximum value around $N_\tau=28$.  }\label{Fig:b61mD}
\end{figure}

On the other hand the Gauss-law approach is based on the assumption of a non-perturbative bound state immersed in a weakly coupled plasma. As we could expect, around  $T_C$ the values of $\Im V$ show sizable deviations from the Gauss-law prediction, which is probably related to the fact that the HTL permittivity is incapable of capturing the relevant physics close to the transition. The best fit values for $m_D$, as well as the uncertainty including the propagated errors from the estimation of the vacuum parameters are given in Tab.~\ref{Tab:DebyeMassSU3-Coarse} and plotted in Fig.~\ref{Fig:b61mD}

\begin{table*}
\small
\begin{tabularx}{17cm}{ | c | X | X | X | X | X | X | X | X | X |}
\hline
	$T/T_C$ & 1.4 & 1.27 & 1.67 & 1.08 & 1. & 0.93 & 0.88 & 0.78 & 0.39 \\ \hline
	$m_D$  [MeV] & $ 603\pm24$ & $430\pm42$ & $367\pm39$ & $273\pm72$ & $150\pm16$ & $14\pm8$ & $0\pm10$ & $0\pm8$ & $0$\\ \hline	
\end{tabularx}\caption{Debye mass parameter extracted from the generalized Gauss-law fit to the in medium $\Re[V]$ at $\beta=6.1, \xi_b=3.2108$. }\label{Tab:DebyeMassSU3-Coarse}
\end{table*}

Let us take a look the temperature dependence of $m_D/T$ in Fig.~\ref{Fig:b61mD}. As expected from the naive inspection of $\Re[V]$ by eye, the Debye mass parameter takes on non-zero values only above the phase transition located at $T_{C}^{\beta=6.1}=290$MeV. Between $T_C$ and $1.4T_C$, our highest temperature, $m_D/T$ changes between $m_D/T\approx0.5\ldots1.5$. This behavior is quite different from the one of finite and almost constant $m_D$ below $T_C$, we observed previously on finer lattices with smaller volumes, which will be discussed and its validity reassessed in the next section \ref{sec:FineLat}. Instead Fig.~\ref{Fig:b61mD} resembles at least qualitatively the outcome of a study of the potential in dynamical QCD with $N_f=2+1$ flavors based on the ASQTAD action at rather heavy pion masses of $m_\pi=300$MeV. I.e. as long as $T/T_C<m_\pi^{\rm lat}/T_C$ and the quark degrees of freedom are not yet thermally activated, such a similar behavior of $m_D$ is fathomable. 

The values of $m_D$ we find here are systematically lower than those reported from fitting Yukawa-type ansaetze to the large distance behavior of the color singlet free energies on the lattice \cite{Kaczmarek:2005ui}. Differences between the potential and $F^{(1)}$ are discussed later in this section. Note also as crosscheck for the scale setting that we have plotted as gray dashed curve the normalized autocorrelations in Monte-Carlo time of a representative Wilson line correlator. Close to the phase transition we find a sharp peak, whose position agrees with the value of the critical temperature given in \cite{Matsufuru:2001cp}.

First we wish to quantify how strongly this lattice $m_D$ differs from the predictions of pure resummed hard-thermal loop perturbation theory, which by itself actually predicts an upward trend of the ratio $m_D/T$ when approaching the transition temperature from above. A comparison to a continuum computation is only meaningful if a continuum extrapolation of the lattice data is available, which even though it is work in progress has not yet been completed. In the meantime we propose to account for finite lattice spacing artifacts, i.e.\ the higher transition temperature and a non-continuum string tension, by considering the following continuum corrected Debye mass
\beq
m_D^{\rm cont.\,corr.}(t=T/T_C^{\rm phys})=\frac{m_D\left(t\right)}{\sqrt{\sigma(\beta)}}\sqrt{\sigma^{\rm cont}}\label{mDcont},
\eeq
where we use $T_C^{\rm phys}=0.271$GeV and $\sigma_{\rm cont}=0.173$GeV. The former corresponds to the continuum transition temperature of SU(3) Yang-Mills theory and the latter is obtained from a Cornell potential analysis of the Bottomonium bound states as listed in the PDG. 

Furthermore the authors of ref.~\cite{Arnold:1995bh} showed that in perturbation theory the Debye mass can be computed up to leading order, including logarithmic correction up to next to leading order. Beyond this point the magnetic sector of QCD contributes non-perturbatively to the value of $m_D$, which we may attempt to capture by introducing additional terms $\kappa_1$ and $\kappa_2$ that need to be determined numerically

\begin{widetext}
\begin{eqnarray}
m_D^{\rm fit}&=&\underbracket{T g(\mu)\sqrt{\frac{N_c}{3}+\frac{N_f}{6}}}_{LO} \notag
	+\underbracket{\frac{N_c T g(\mu )^2}{4 \pi } \log
   \left(\frac{\sqrt{\frac{N_c}{3}+\frac{N_f}{6}}}{g(\mu )}\right)}_{NLO} +\underbracket{\kappa_1 \,T g(\mu )^2+\kappa_2\, T g(\mu )^3}_{\rm non\, pert.}. \label{Eq:mD}
\end{eqnarray}
\end{widetext}

This formula reduces at high temperature to the well known HTL NLO result due to asymptotic freedom, while at lower temperatures the presence of $\kappa_1$ and $\kappa_2$ allow for a deviation from the purely perturbative behavior. Thus the parameters $\kappa_1,\kappa_2$ should in principle be extracted from the high temperature behavior of the Debye mass, were the HTL expansion is expected to be closest to the full physics. 

Here we evaluate the above expression at the scale $\mu=\pi T$, using for the running of the coupling a four loop computation of $g(\mu)$ with a $\Lambda_{\rm QCD}=0.2145$GeV \cite{Vermaseren:1997fq} appropriate if the renormalization group flow is initiated at a scale with $N_f=5$ active flavors. As shown by the red solid line in Fig.~\ref{Fig:b61mD} this fitting ansatz for the Debye mass allows us to reproduce our values of $m_D$ within errors.  We obtain finite values for the non-perturbative parameters 
\beq
\kappa_1=1.97 \pm 0.44, \quad \kappa_2 =-1.09\pm0.26,
\eeq
which indicate that the behavior observed on the lattice, in particular the downward trend towards $T=T_C$ requires significant deviation from the purely perturbative predictions. Note that our fit can contain significant systematic errors due to the fact that the HTL expansion might not be a precise description of the plasma close to $T_C$.

\begin{figure}
\centering
\begin{overpic}[scale=0.4,trim=0cm 1.8cm 0cm 1.8cm, clip=true]{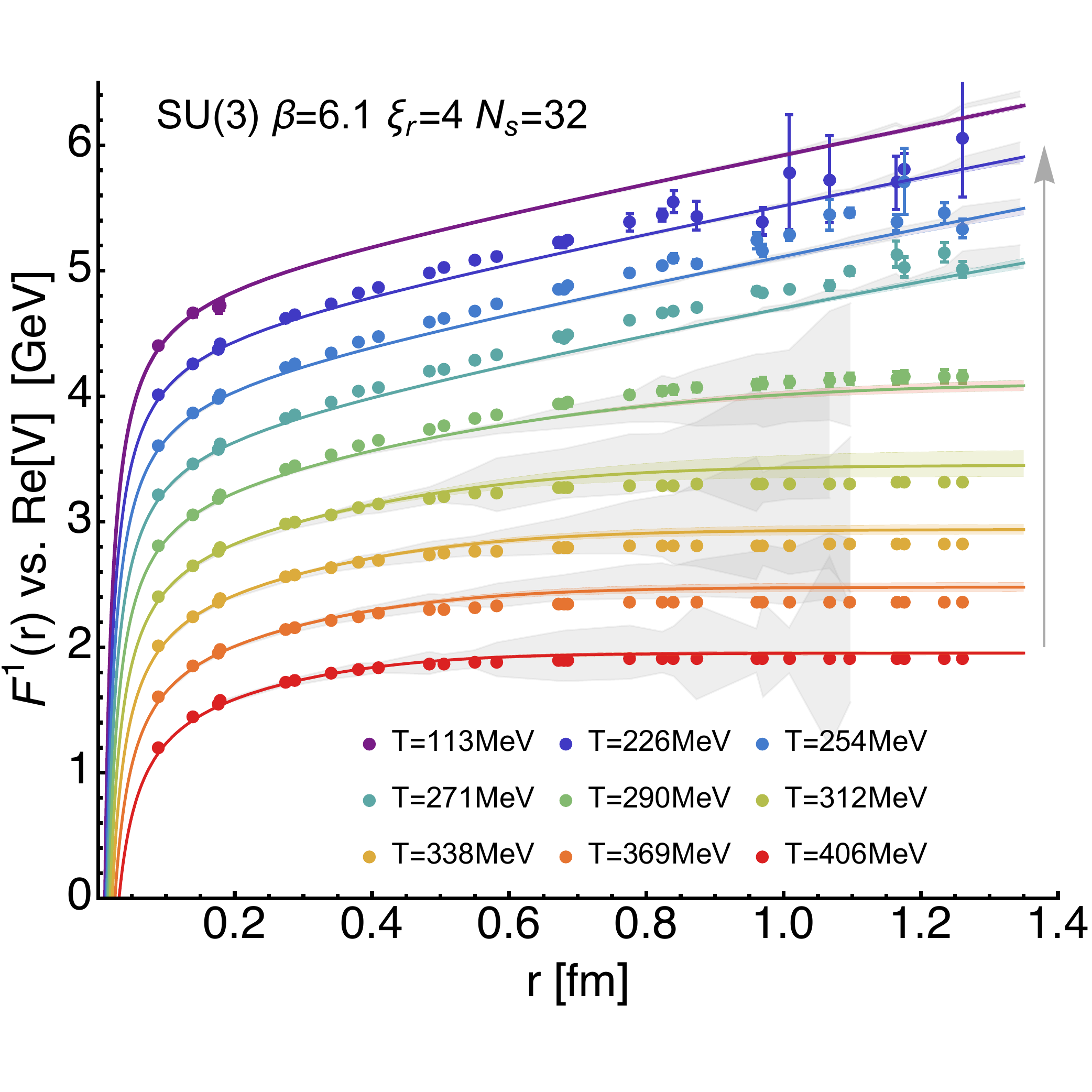}
\put (96,45) {\begin{turn}{90} \color{mygray}\tiny manually shifted \end{turn}}
\end{overpic}
\caption{Comparison of the color singlet free energies (colored dots) to the analytic parametrization of the in-medium heavy quark potential (solid lines). The values are shifted by hand in y for better readability, as indicated by the gray arrow. The gray errorbands represent the systematic errors of the lattice values of Re[V] and are shown to better assess the uncertainties of its Gauss-law parametrization.}\label{Fig:b61F1}
\end{figure}

Now that we have investigated the differences between our values for the Debye mass parameter from the Gauss-law ansatz and $m_D$ in resummed perturbation theory, we proceed to a comparison with another popular definition from the large distance behavior of the color singlet free energies. The values reported in the literature based on a fit of a Yukawa type ansatz are systematically larger than what we observe. One reason for this discrepancy lies in the quantitatively and qualitatively different behavior of $\Re[V]$ and $F^{(1)}$, as can been seen in Fig.~\ref{Fig:b61F1}. The colored points denote the color singlet free energies, while the solid lines correspond to the Gauss-law fit, with which we reproduced our values for $\Re[V]$. The gray errorbands correspond to the systematic uncertainty of the lattice $\Re[V]$ and are shown to better assess the systematic uncertainty in the fit.

It is known that at $T=0$ the color singlet free energies agree with the purely real vacuum potential, in particular they show the same string tension. Consistently with what has been reported in the literature (also for dynamical QCD \cite{Kaczmarek:1900zz}) we find that at finite temperature but below the deconfinement transition, $F^{(1)}$ actually exhibits a stronger string tension than in vacuum. From the point of view of static inter-quark interactions this is counterintuitive, since thermal fluctuations are expected to at most weaken the binding properties, if they influence them at all. Therefore one might need to reconsider in how far $F^{(1)}$ can help us to quantitatively shed light on the in-medium modification of static quark binding in medium. 

Around $T_C$, $F^{(1)}$  then seems to transition to a screened behavior much more quickly than $\Re[V]$ and only at high temperatures $T\approx 1.5T_C$ the two begin to approach each other again, with $F^{(1)}<\Re[V]$ systematically. If we had carried out a determination of the Debye mass parameter using the color singlet free energies, we would have started off at low temperatures from a larger than vacuum string tension and would have observed a stronger apparent screening shortly above $T_C$, both factors leading to a larger value of a screening mass around the phase transition extracted in that way. 

Note that the proper in-medium potential $\Re[V]$ on the other hand never shows a stronger than vacuum string tension below $T_C$, in fact it seems to be rather unaffected by thermal fluctuations below $T_C$. This is expected from the fact that the only color neutral objects capable of disturbing the string between the static quarks are glueballs, which due to their large mass are highly suppressed at the temperatures simulated here.

\begin{figure}[t!]
\centering
\includegraphics[scale=0.42,trim=2.2cm 0cm 0cm 0cm, clip=true]{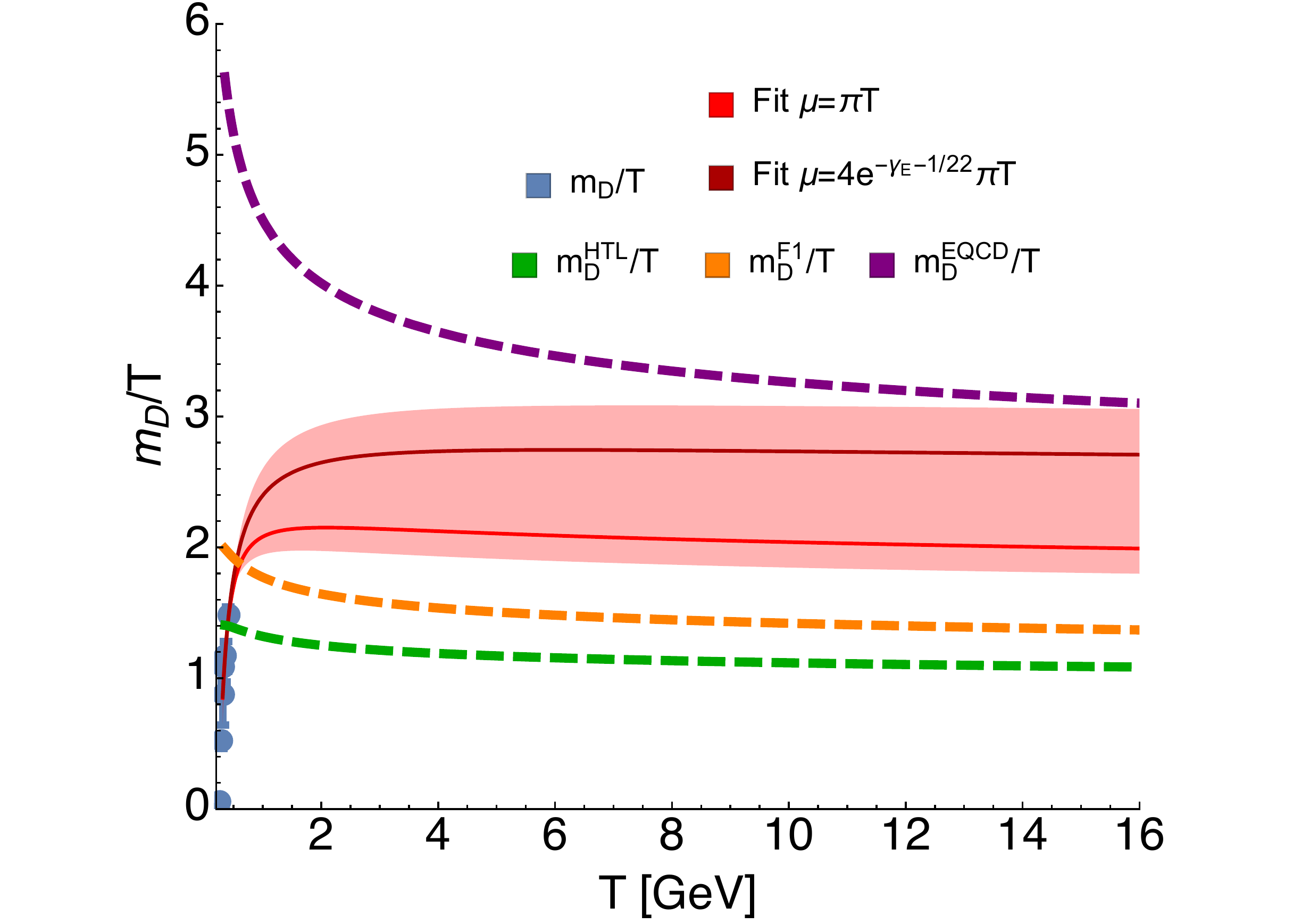}
\caption{Extrapolation to high $T\gg T_C$ of our fits to the Debye mass (red, solid) for two different choices of scale setting, i.e.\ $\mu=\pi T$ and $\mu_{\rm EQCD}=4e^{-\gamma_E-1/22}\pi T$ used in the EQCD study ref.~\cite{Kajantie:1997pd}. While the fits are both able to reproduce our values of $m_D/T$ close to $T_C$ a sizable uncertainty in extrapolating to higher temperatures exists denoted by the red errorband. We find that for $T\gg T_C$ the fits lie systematically above the values obtained from fitting $m_D/T$ from the color singlet energies (orange dashed) from ref.~ \cite{Kaczmarek:2005ui}. When using the same scale  $\mu_{\rm EQCD}$ as in the EQCD study, our extrapolation seems to become compatible to their values from around $T>16$GeV. The pure HTL behavior is given as green dashed curve.}\label{Fig:b61mDvsLit}
\end{figure}

Interestingly studies of dimensionally reduced QCD (EQCD) have also found that the electric screening lies far above the perturbative values \cite{Kajantie:1997pd}. Since such computations are expected to be valid at $T\gg T_C$, we can only compare them to an extrapolation of our fit, as shown in Fig.~\ref{Fig:b61mDvsLit}. Using the scale $\mu=\pi T$, the high temperature extrapolation remains below the values obtained for EQCD, however when using a fit based on the same scale $\mu_{\rm EQCD}=4e^{-\gamma_E-1/22}\pi T$, where $\gamma_E$ denotes Euler's constant, we find that the large $T$ behavior becomes compatible above $T=16$GeV. On the other hand $\mu_{\rm EQCD}$ corresponds to a rather large choice, since from thermodynamics one would expect the characteristic scale of two-particle interactions to be $2\cdot \frac{3}{2} T=3T\approx \pi T$. Correspondingly the fit values for the $\kappa$ parameters in that case also become unnaturally large $\kappa^{\mu_{\rm EQCD}}_1=4.8\pm1.1$ and $\kappa^{\mu_{\rm EQCD}}_2=-3.2\pm0.7$. We are reminded that an evaluation of the Debye mass parameter at high temperatures is urgently needed to reduce the uncertainty from extrapolations.

\begin{figure}
\centering
\includegraphics[scale=0.4,trim=0cm 0cm 0cm 0cm, clip=true]{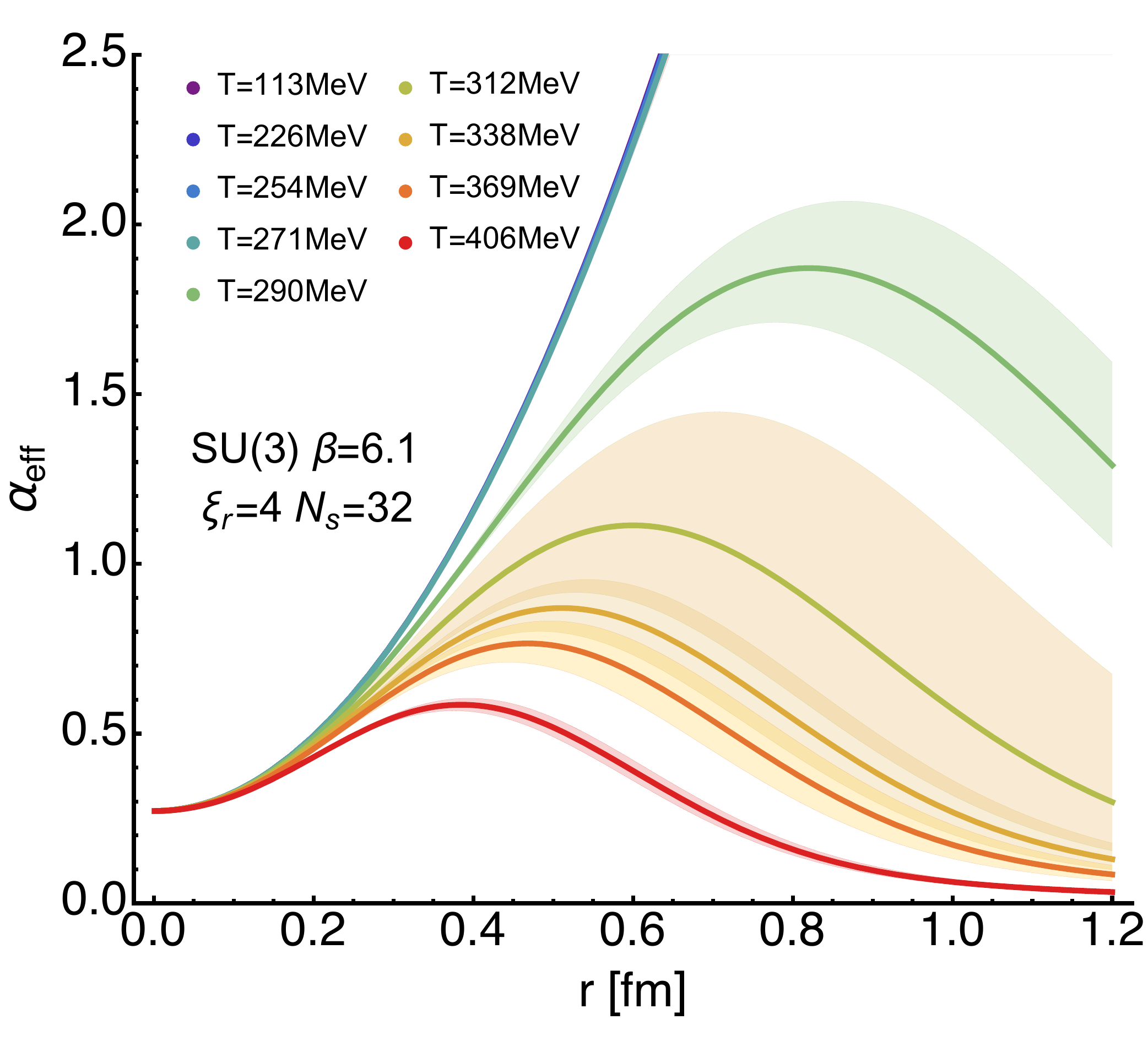}
\caption{Effective coupling $\alpha_{\rm eff}$ from the derivative of the real part of the potential $\alpha_{\rm eff} = r^2\partial_r\Re[V]$. For the values shown here the Gauss-law fit is used to compute the derivative. We see that the residual influence of the confining potential already sets in at $r\approx0.2$fm and remains visible even at the highest temperature investigated.}\label{Fig:b61Aeff}
\end{figure}

As last point of our investigation of the $\beta=6.1$ ensembles we turn to the effective coupling, defined as
\beq
  \alpha_{\rm eff}(r,T)=r^2\partial_r\Re[V](r,T),
\eeq
which allows us to e.g.\ determine the residual influence of the vacuum string tension at finite temperature. In a purely Coulombic potential $\alpha_{\rm eff}$ would be constant and reproduce $\alpha_S$. In the presence of a running coupling, at small distances a deviation from a constant would be observed. On the other hand, for a confining potential $\alpha_{\rm eff}$ exhibits a quadratic behavior. Screening, i.e.\ a flattening off of the potential in turn leads to a vanishing of the effective coupling. In Fig.~\ref{Fig:b61Aeff} we show its values computed from taking the derivative of the corresponding Gauss-law fit to $\Re[V]$ down to the smallest available distance on the lattice $r\approx0.1$fm.

We find that at small $r$, $\alpha_{\rm eff}$ approaches $\alpha_S$, while it has not yet asymptoted to a constant value. In the absence of a running coupling in the Gauss-law ansatz, the reason lies in that the linear confining rise and its remnants at $T>0$ still contributing to $\alpha_{\rm eff}$ even at $r\approx0.1$fm. Below $T_C$  we find a quadratic rise at larger distances and as expected from the confining nature of the potential. It is only in the deconfined phase that a turnover point appears, due to the onset of screening. This is fully consistent with the values of the Debye mass parameter extracted above and in particular reiterates that $\Re[V]$ never shows a stronger string tension than in vacuum.

\begin{figure}
\centering
\includegraphics[scale=0.4, trim=0cm 2cm 0.1cm 2cm, clip=true]{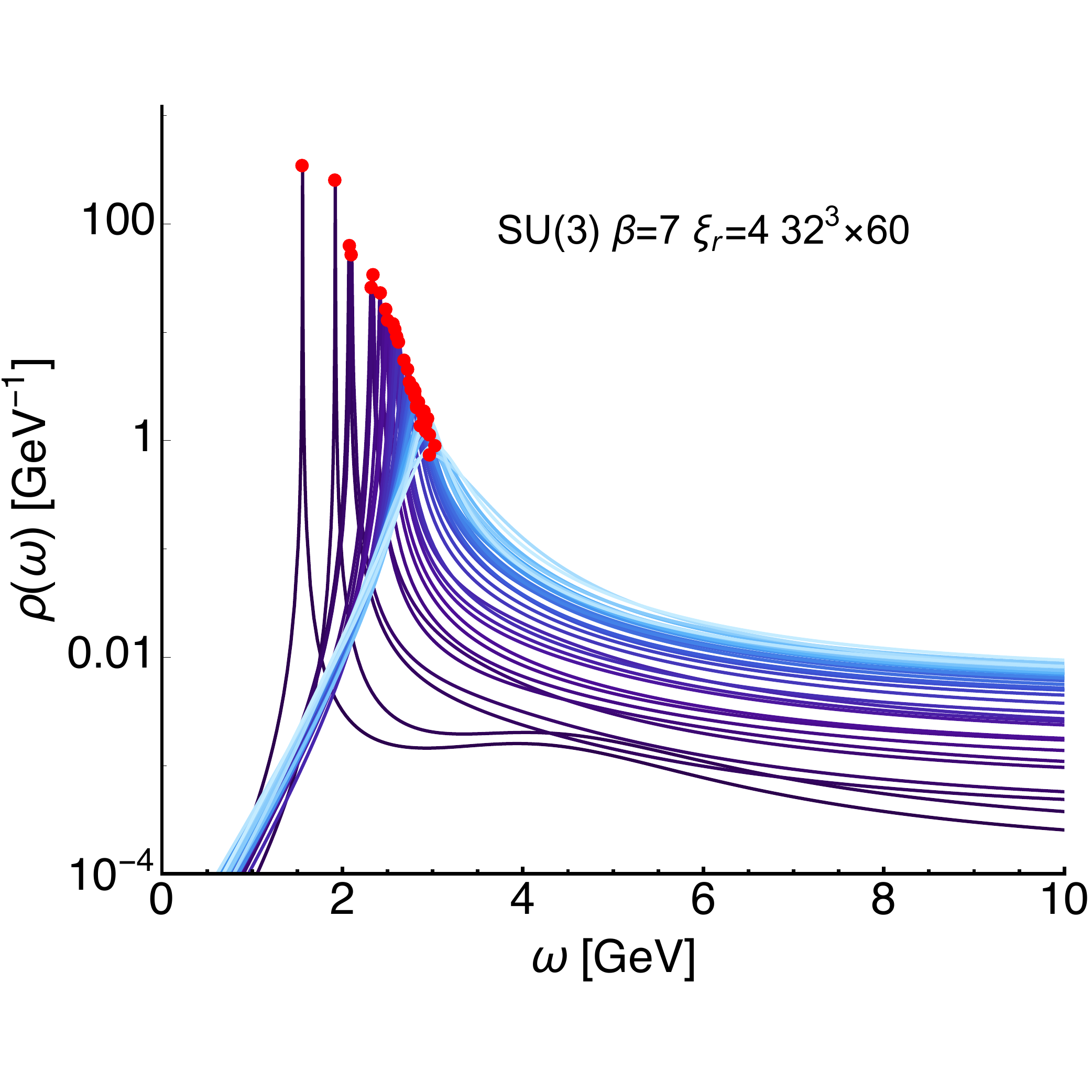}
\includegraphics[scale=0.4, trim=0cm 2cm 0.1cm 2cm, clip=true]{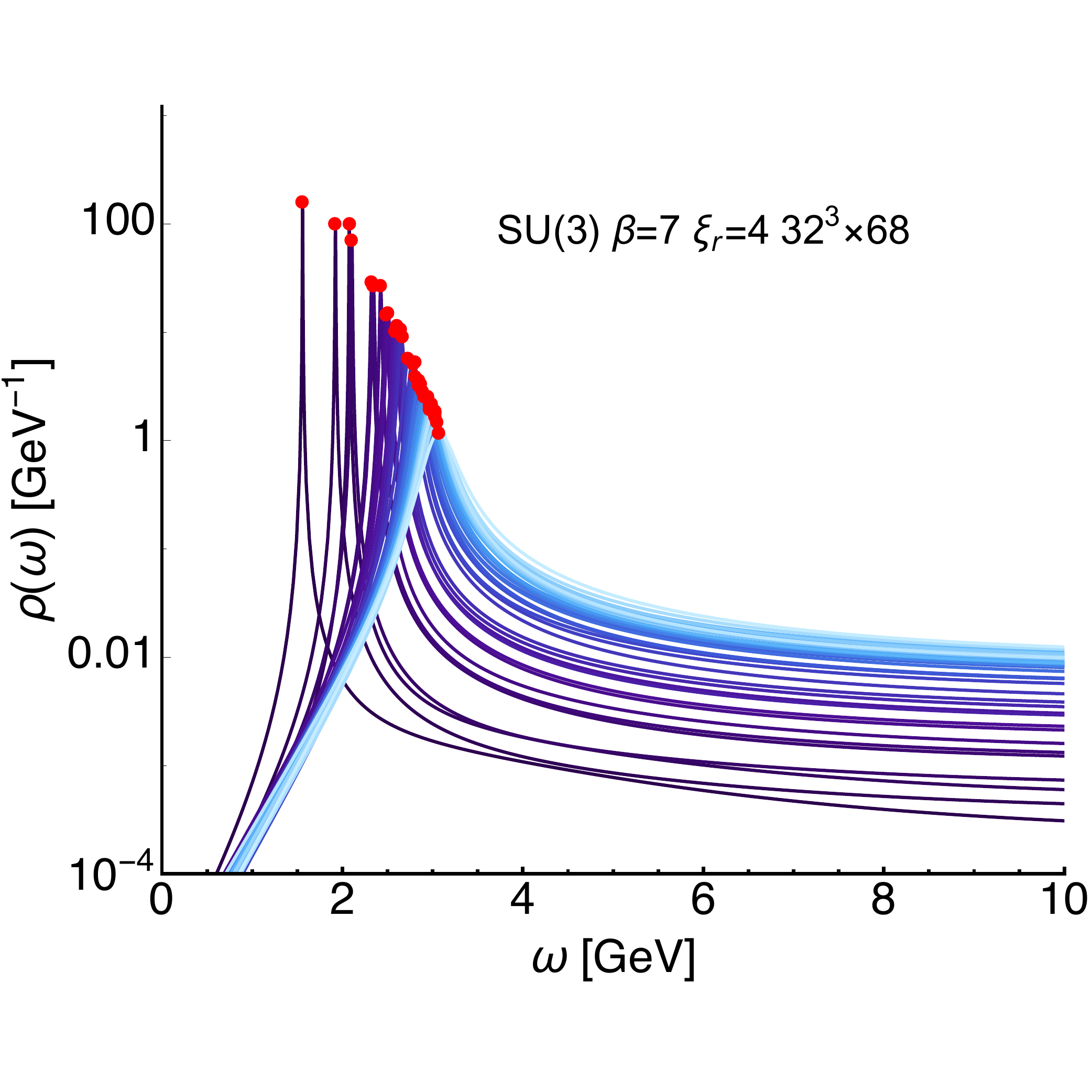}
\caption{Spectral reconstructions for two of the new ensembles at $N_\tau=60$ (left) and $N_\tau=68$ (right).}\label{Fig:b7SpecRecs}
\end{figure}

\section{Finte volume artifacts on $\beta=7$ lattices at $L=1.25{\rm fm}$}
\label{sec:FineLat}

\begin{table*}
\small
\begin{tabularx}{17cm}{ | c | X | X | X | X | X | X | X | X | X | X | X | X |}
\hline
	SU(3): $N_\tau$ & 24 & 32 & 40 & 48 & 56 &60 & 64 & 68 & 72 & 80 & 96 & 192\\ \hline
	$T$[MeV] & 839 & 629 & 503 & 419 & 360 & 335 &  315 & 296 & 280 & 252 & 210 & 105\\ \hline
	$T/T_C$ & 3.11 & 2.33 & 1.86 & 1.55 & 1.33 & 1.24 &  1.17 & 1.10 & 1.04 & 0.93 & 0.78 & 0.39\\ \hline
	$N_{\rm meas}$ & 3270 & 2030 & 1940 & 1110 & 1410 & 2420 & 1530 & 2190 & 860 & 1190 & 1800 & 900\\ \hline
\end{tabularx}\caption{Quenched lattice ensembles at $\beta=7,\xi_b=3.5, N_s=32$ corresponding to a physical $a=0.039{\rm fm}$ and $L=1.25{\rm fm}$. }\label{Tab:LatParmb7}
\end{table*}

In this section we return to a parameter set we had previously used to study the in-medium modification of the inter-quark potential \cite{Burnier:2014ssa,Rothkopf:2011db} and which has been deployed in the literature e.g.\ for the direct extraction of Charmonium spectral functions from current-current correlators \cite{Asakawa:2003re}. For these anisotropic lattices we have  $\beta=7$ and a bare anisotropy of $\chi_b=3.5$, which lead to a physical lattice spacing of $a_s=0.039=4a_\tau$ according to \cite{Asakawa:2003re}. Similar to the choice in the literature $N_s=32$ points along the spatial axes, i.e.\ a box size of $L=1.25$fm is deployed here. What we will argue in the following is that this small box size and the correspondingly high infrared cutoff introduce an artificial screening behavior in the real-part of the in-medium potential, which even persists at low temperatures and in turn impedes also the determination of the physical lattice spacing using e.g.\ the Sommer scale related to the $T=0$ string tension.

The temperatures around the deconfinement transition $T_{C}^{\beta=7}=270(10)$MeV are probed in the range $T=3.11-0.78T_C$ corresponding to  $N_\tau=24-96$. For the low temperature reference at $T=0.39T_C$ we simulated lattices with $N_\tau=192$. In addition to the previously analyzed ensembles from Refs. \cite{Burnier:2014ssa,Burnier:2015nsa} we both add here two high statistics ones close to $T_C$ at $N_\tau=60$ and $N_\tau=68$, i.e.\ $T=1.24T_C$ and $T=1.10T_C$ respectively and replace the previous configurations at $N_\tau=64$ with a larger number of newly generated ones. In particular we have made sure that these new ensembles are well thermalized with a long burn in period of 120000 sweeps.  A summary is provided in Tab.~\ref{Tab:LatParmb7}.

Compared to the coarser lattices of sec.\ref{sec:CoarseLat}, the naive UV cutoff $\sqrt{3}\pi/a_s$ corresponds to $\Lambda_{\rm UV}=27.5$GeV, while the IR cutoff takes on a relatively large value of $\Lambda_{\rm IR}=0.992$GeV. Already at this point we would like to point out that this value of $\Lambda_{\rm IR}$ is much larger than any temperature investigated here, as well as the intrinsic scale of QCD $\Lambda_{\rm QCD}/\Lambda_{\rm IR}\ll1$ that controls the physics of confinement. It is known that the effects of an infrared cutoff manifest themselves in both an artificial mass for the dynamical degrees of freedom in a simulation, as well as in the complete absence of modes with wavelength larger than its inverse. This fact has to be kept in mind when interpreting and assessing the outcome of the potential extraction.

\begin{figure}
\centering
\begin{overpic}[scale=0.38,trim=0cm 1.cm 0cm 0.8cm, clip=true]{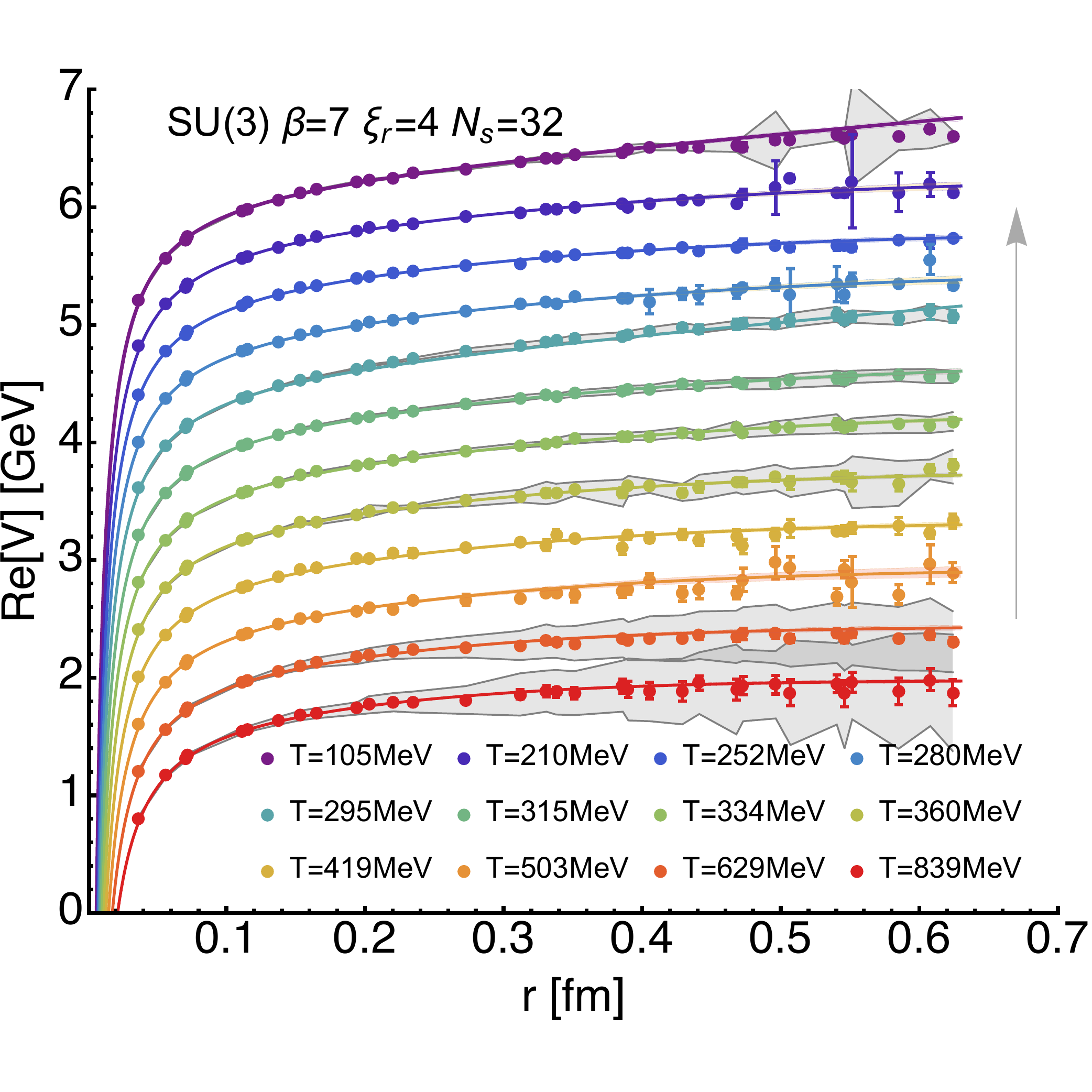}
\put (94,45) {\begin{turn}{90} \color{mygray}\tiny manually shifted \end{turn}}
\end{overpic}
\begin{overpic}[scale=0.19,trim=0cm 0cm 0cm 0cm, clip=true]{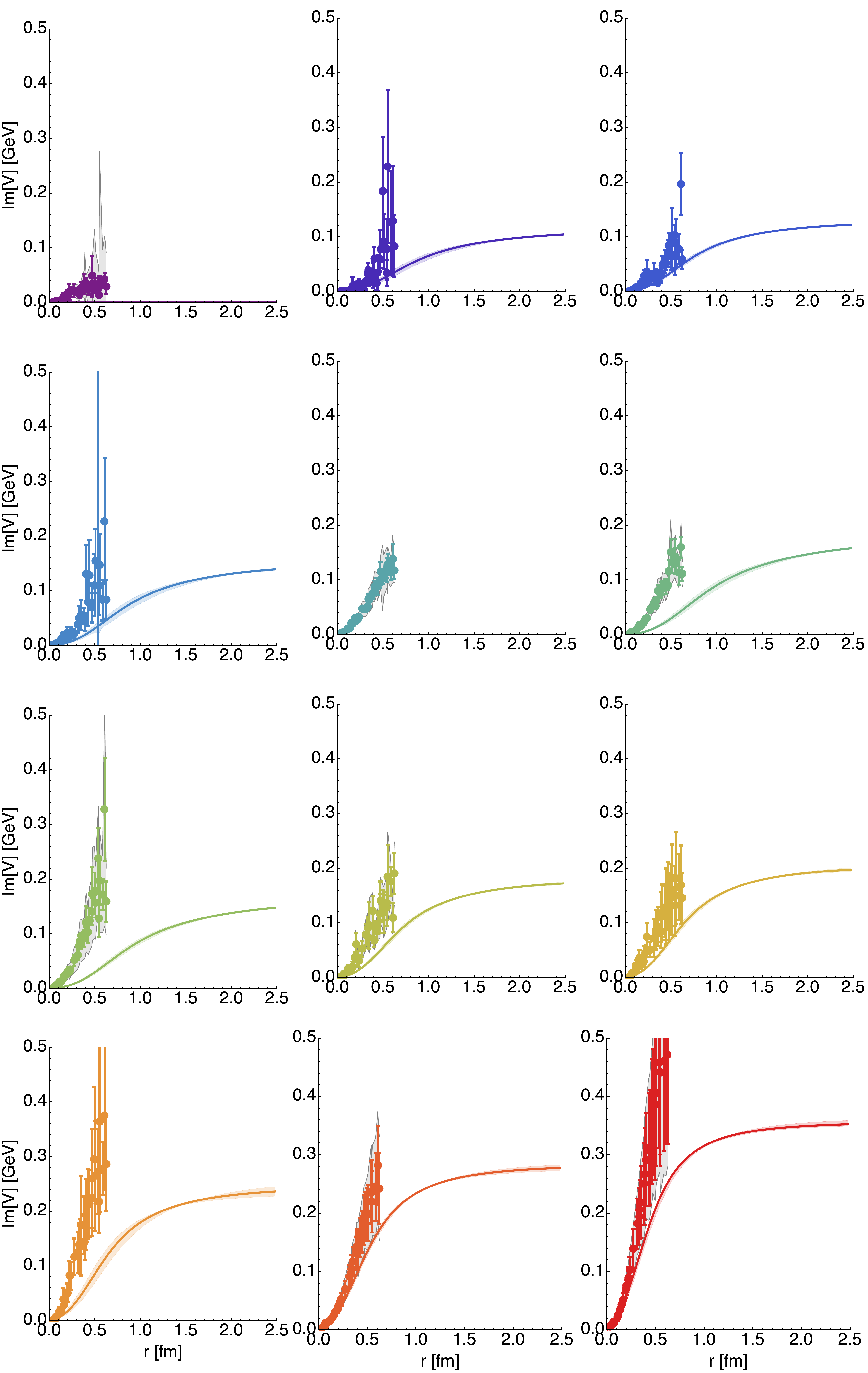}
 \put (51,6) {\tiny$T=839$MeV}  \put (30,6) {\tiny$T=629$MeV}  \put (10,6) {\tiny$T=503$MeV}
 \put (51,45) {\tiny$T=419$MeV}  \put (30,45) {\tiny$T=360$MeV}  \put (10,45) {\tiny$T=334$MeV}
 \put (51,70) {\tiny$T=315$MeV}  \put (30,70) {\tiny$T=295$MeV}  \put (10,70) {\tiny$T=280$MeV}
 \put (51,95) {\tiny$T=252$MeV}  \put (30,95) {\tiny$T=210$MeV}  \put (10,95) {\tiny$T=105$MeV}
\end{overpic}
\caption{Extracted values of the real (top) and imaginary part (bottom) of the in-medium potential (colored points) at $\beta=7$. The values of $\Re[V]$ are shifted manually in y-direction for better readability as indicated by the gray arrow. The values of $\Im[V]$ are plotted individually as grid from lowest temperature (top left box) to highest temperature (bottom right box). The colored errorbars denote statistical, while the gray errorband systematic uncertainty.}\label{Fig:b7ReVImV}
\end{figure}
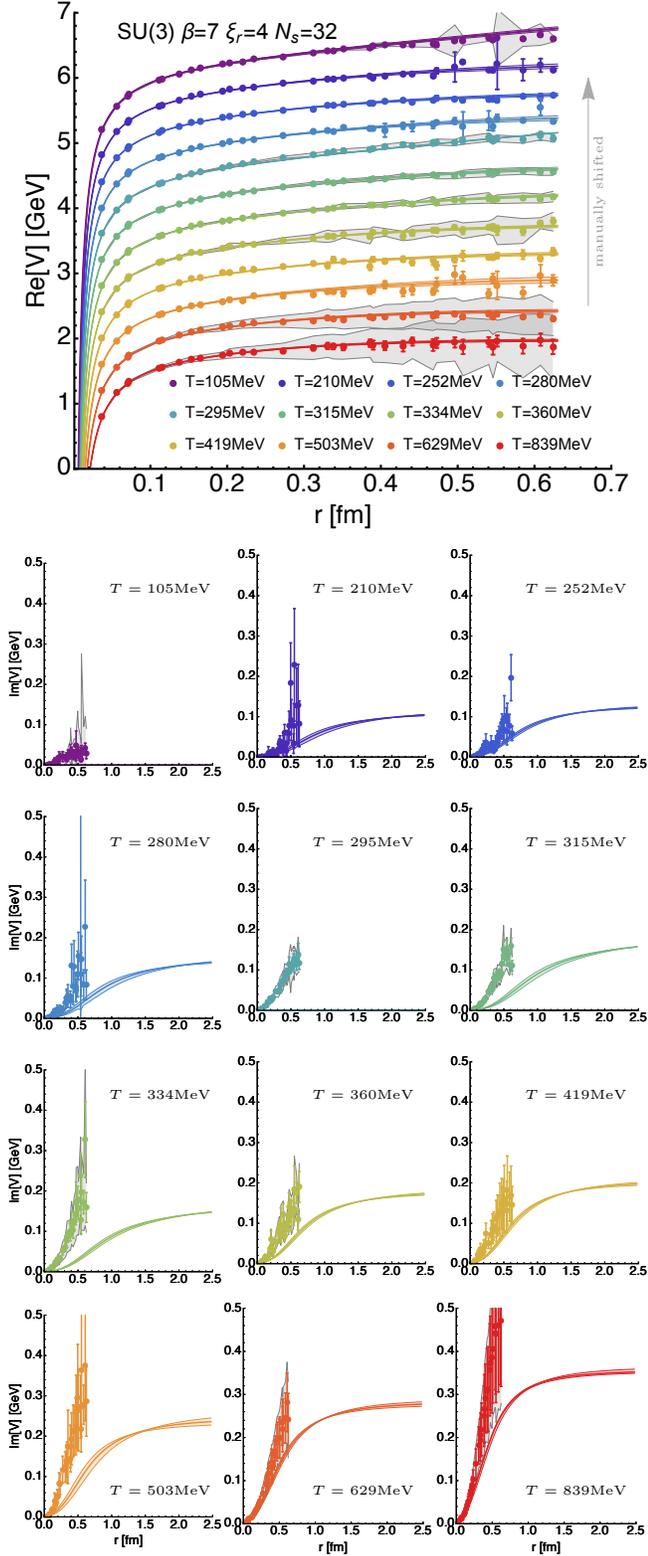

Similar to the coarser lattices we fix iteratively to Coulomb gauge with tolerance $\Delta_{\rm GF}=10^{-15}$ before measuring the Wilson line correlators along the spatial axes, as well as the $\sqrt{2}$ and $\sqrt{3}$ diagonals for all possible separations in Euclidean time direction. These datasets, excluding the first $\tau=0$ and last $\tau=\beta$ point are fed to the spectral reconstruction code based on the BR method. In contrast to the previous analysis carried out in \cite{Burnier:2014ssa,Burnier:2015nsa}, we choose the numerical frequency range here equal to the one deployed for the coarser lattices, i.e.\ $\omega_{\rm num}\in[-5,5]$ and reanalyze all available datasets for consistency. No significant deviation between the previous reconstructions has been found. In Fig.~\ref{Fig:b7SpecRecs} we show two representative sets of Bayesian spectra for all considered values of $r$ on the newly generated ensembles at $N_\tau=60$ (top) and $N_\tau=68$ (bottom). The tip of each of the well pronounced lowest lying peaks is marked by a red circle. Its position and width yields the values of the in-medium potential, which is plotted as colored points in Fig.~\ref{Fig:b7ReVImV}.

We have shifted the values of $\Re[V]$ by hand for better readability in the top panel of Fig.~\ref{Fig:b7ReVImV} as indicated by the gray arrow and plot the statistical errors as colored bars. The systematic errors denoted by the gray errorbands have been determined only for part of the ensembles, in particular however among the newly generated ones $N_\tau=60,64,$ and $68$. We use the variation from changing the default model amplitude by two orders of magnitude, as well as discarding $10\%$ of the small $\tau$ and/or $10\%$ of the large $\tau\approx\beta$ datapoints as an estimate.

Just as in the previous analysis we find that there appears to exist only a gradual change in behavior of $\Re[V]$ with increasing temperature. Due to limited statistics in the older analysis, $N_\tau=64$ seemed to exhibit an anomalously strong linear rise. This effect has vanished after increasing statistics and the slope at $N_\tau=64$ now lies within the trend of neighboring $N_\tau=60$ and $N_\tau=72$.  On the other hand, it is now $N_\tau=68$, which is found to show an almost a vacuum like linear rise. The reason for this outlier however lies in the fact that the $N_\tau=68$ simulations show extremely long autocorrelation times in Monte-Carlo time. In turn even after collecting more than 2100 measurements on individual configuration, the actual statistics are reduced by around a factor ten to one hundred, making this result rather unreliable. Interestingly this issue does not lead to an increase in the systematic errorbars compared to e.g.\ $N_\tau=64$.

\begin{figure}[t!]
\centering
\includegraphics[scale=0.4,trim=3.5cm 0cm 0cm 0cm, clip=true]{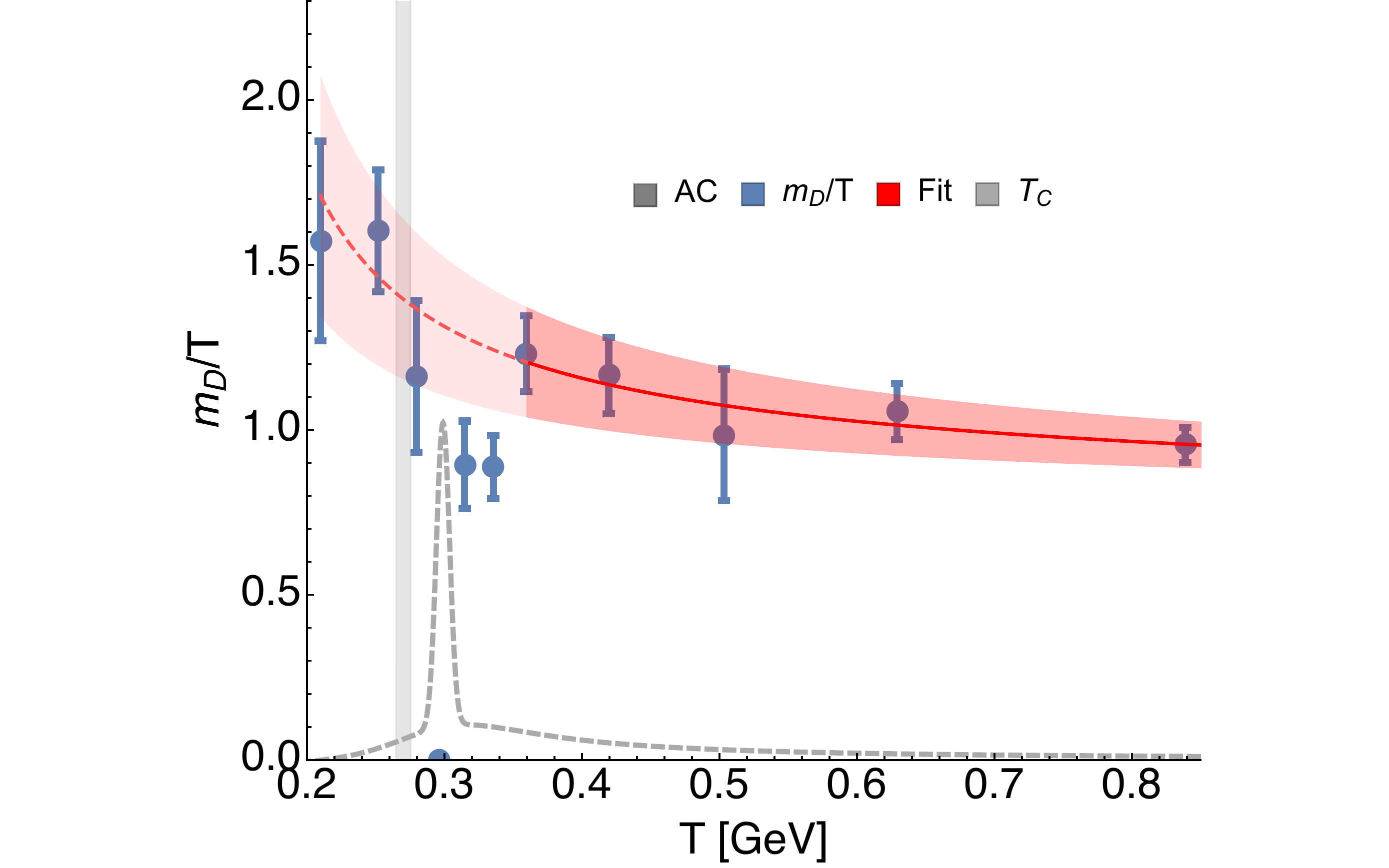}
\caption{Debye mass parameter of the heavy quark potential for $\beta=7$, describing its in-medium modification. The blue points denote the extracted values for $m_D/T$, while the red curve corresponds to a fit based on HTL perturbation theory amended by non-perturbative correction terms. The transition temperature determined via the Polyakov loop is given as a gray band. Complementarily we show the autocorrelation time of the Wilson line correlators at $r/a_s=5.2$ and $\tau/a_\tau=14$ normalized to unity at its maximum value around $N_\tau=68$.  }\label{Fig:b7mD}
\end{figure}

For the sake of completeness we continue towards performing the Gauss-law fit by determining first the vacuum parameters that enter into this ansatz from the lowest temperature result at $N_\tau=192$. Restricting to the region of $r<0.3$fm we obtain
\begin{align}
\nonumber&\alpha_S=0.201\pm0.004,\quad \sigma=0.186\pm0.008{\rm GeV}^2\\ &c=2.58\pm0.01{\rm GeV}
\end{align}
where the errors are again estimated from a variation of the upper and lower end of the fitting range by six steps each. These agree within errors with our previous values published in \cite{Burnier:2015nsa}. The Debye mass parameter we find to fit the in-medium values of $\Re[V]$ at $\beta=7$ best are compiled in Tab.~\ref{Tab:DebyeMassb7} and plotted in Fig.~\ref{Fig:b7mD}. In addition we also perform a fit to the extracted values based on Eq.\eqref{Eq:mD}, which yields as best fit parameters
\beq
\kappa_1=-0.67 \pm 0.06, \quad \kappa_2 =0.34\pm0.06
\eeq
and which is plotted as solid line in Fig.~\ref{Fig:b7mD}. The uncertainties in the $\kappa$ values arise from the error in estimating $m_D$ itself.

\begin{table*}
\small
\begin{tabularx}{17cm}{ | c | X | X | c | X | X | X | X | X | X | X | X | X |}
\hline
	$T/T_C$ & 3.11 & 2.33 & 1.86 & 1.55 & 1.33 & 1.24 &  1.17 & 1.10 & 1.04 & 0.93 & 0.78 & 0.39\\ \hline
	$m_D {\rm [MeV]}$ & $ 800\pm45$ & $664\pm53$ & $496\pm100$ & $489\pm48$ & $442\pm41$ & $298\pm 32$ & $281\pm42$ & $\approx 0$ & $325\pm64$ & $403\pm46$ & $330\pm63$ & 0\\ \hline	
\end{tabularx}\caption{Debye masses extracted from the generalized Gauss-law fit to the in medium $\Re[V]$ at $\beta=7, \xi_b=3.5$.}\label{Tab:DebyeMassb7}
\end{table*}

Let us discuss in detail the temperature dependence of the Debye mass parameter in Fig.~\ref{Fig:b7mD}. It significantly differs from the two qualitatively similar results we obtained on the coarser lattices at $\beta=6.1$ presented in sec.\ref{sec:CoarseLat}, as well as in dynamical $N_f=2+1$ QCD with heavy pions. Instead of decreasing in value towards zero  below the phase transition $m_D$ here appears to become constant within the relatively large errors. I.e. for $m_D/T$ an upward trend ensues towards lower temperatures. The only exception is the single outlier at $N_\tau=68$, which shows essentially vanishing Debye mass. Deceptively, the comparatively small values of $\kappa_1$ and $\kappa_2$ if taken at face value would suggest that the behavior found here would actually resemble the predictions from HTL perturbation theory.

\begin{figure}[t!]
\centering
\includegraphics[scale=0.4,trim=0cm 0cm 0cm 0cm, clip=true]{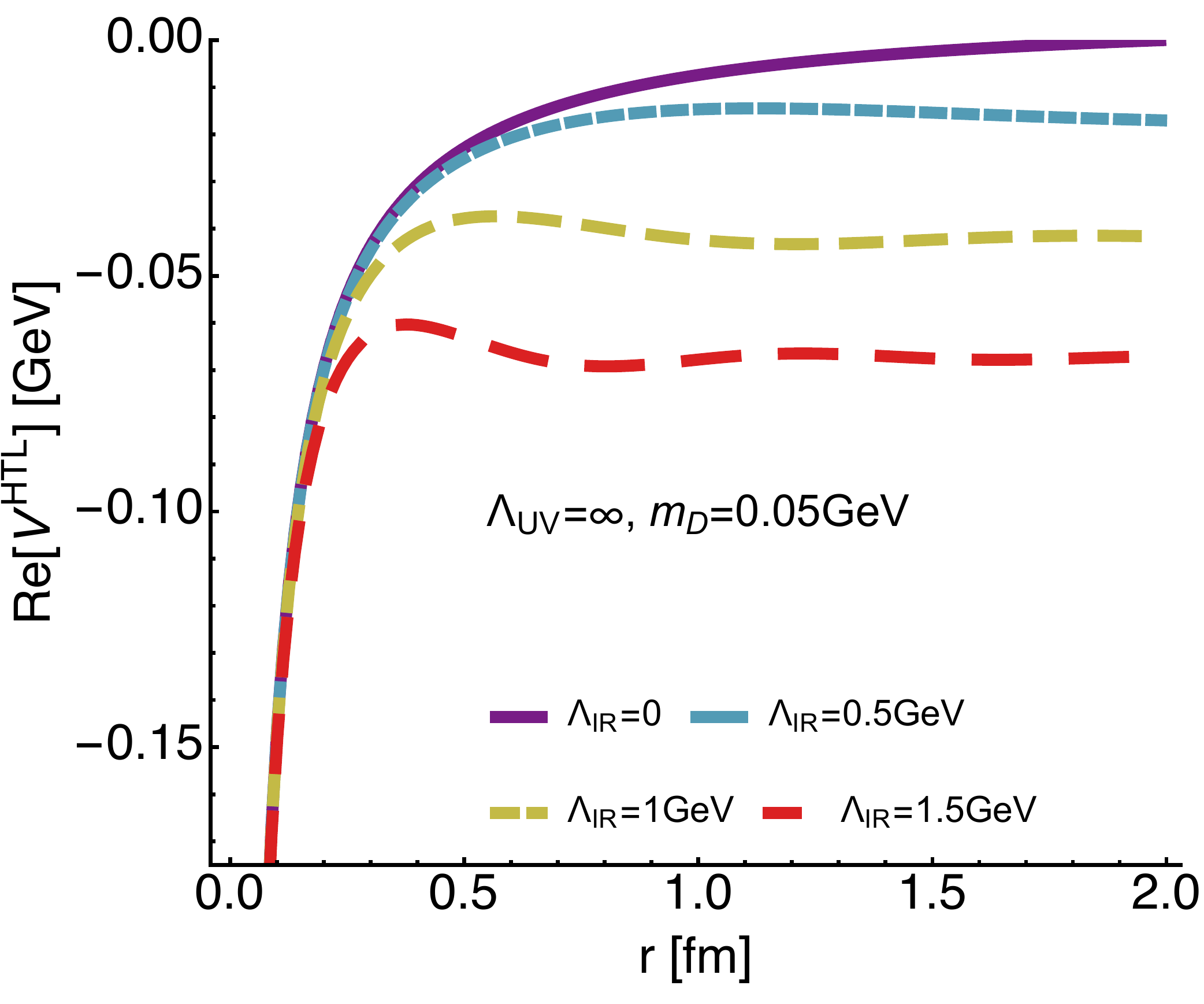}
\caption{The real-part of the in-medium potential in resummed hard-thermal loop perturbation theory in the absence (solid curve) and presence of an infrared cutoff (dashed curves).  }\label{Fig:b61CutOff}
\end{figure}

After generating and analyzing the three new ensembles for $\beta=7$, we believe that this is not the case and instead the finite box size has rendered our results close to and below $T_C$ unreliable. By that we mean the following: it is known that the presence of an infrared cutoff, which estimated naively here with $\Lambda_{\rm IR}=0.992$GeV takes on a comparatively large value, can lead to artifacts in numerical simulations that mimic the effects of a mass for the dynamical degrees of freedom. As an example let us see how in resummed perturbation theory the values of $\Re[V]$ react to the introduction of spherical cutoffs
\beq
\Re[V^{\rm HTL}](r)=- \int_{\Lambda_{\rm IR}}^{\Lambda_{\rm UV}}  \,dk\,\frac{4\pi}{(2\pi)^3}\frac{k^2}{k^2+m_D^2}\big(\frac{sin[kr]}{kr}-1\big).
\eeq
In Fig.~\ref{Fig:b61CutOff} we show the above formula evaluated for a arbitrary choice of $m_D=0.05$GeV and a $\Lambda_{\rm UV}=\infty$. And indeed the main effect that a finite $\Lambda_{\rm IR}$ results in is a flattening off of $\Re[V]$ at earlier distances (dashed lines) than in the thermodynamic limit (solid line), which mimics to a certain degree physical screening. Thus we learn that a too small box can lead to behavior similar to the presence of an unnaturally large Debye mass.

It is in light of this fact that we now reassess the behavior of $\Re[V]$ in Fig.~\ref{Fig:b7ReVImV} and $m_D$ in Fig.~\ref{Fig:b7mD}. While at higher temperatures the comparatively large Debye mass acts as physical cutoff itself, it at some point will become smaller than the artificial mass introduced by $\Lambda_{\rm IR}$ when approaching the phase transition from above. At that point the apparent screening mass of the potential will become constant as it is dominated by the effects of the small box. It is exactly such a constant behavior of the Debye mass around $T_C$ within statistical errors which we find here\footnote{The fact that $N_\tau=68$ does not follow this behavior is probably related to the long autocorrelation times occurring at this number of axis points, which deteriorates the reliability of the spectral reconstruction}.

We believe that the large infrared cutoff has furthermore impeded the setting of the physical scale and thus the determination of the transition temperature for these lattices. If the Sommer scale is used to determine $a_s$ in physical units, one needs to identify the string tension of the $T=0$ potential to its physical values obtained from a potential model analysis. If however even at low temperatures an artificial screening is present in the system this identification will lead to an incorrect assignment. An indication of a possible mismatch of scales lies in that the maximum of the normalized autocorrelation time in the Wilson line correlators occurs here at $N_\tau=68$, which however is identified with a temperature around $10\%$ above $T_C$. In case of $\beta=6.1$, where a reliable high precision scale setting was available, $T_C$ and the maximum in autocorrelation time consistently agreed.

We therefore believe that it is paramount to repeat the scale setting procedure for the $\beta=7$ parameter set using lattices with at least twice physical axis length in spatial direction, which is work in progress. In particular if we wish to reliably investigate the continuum limit of the potential, for which the $\beta=7$ parameter set was originally destined, we have to take make sure to remain safely close to the thermodynamic limit.

\section{Conclusion}
\label{sec:Conclusion}

We have presented both a new analysis of the in-medium inter-quark potential in quenched QCD on ensembles closer to the thermodynamic limit, as well as a reassessment of our previous findings on finer but smaller lattices. 

On the lattices with larger physical volume $\beta=6.1$ we find a distinct behavior of $\Re[V]$  below and above the critical temperature. In the confined phase we see essentially no modification of the potential, in particular the string tension of the linear confining rise remains unaffected by thermal fluctuations. Just above deconfinement on the other hand a clear flattening off of the real-part can be observed, a sign of screening due to liberated color degrees of freedom. Consistently the values of the imaginary part, after taking systematic uncertainties into account, also only show values away from zero at $T>T_C$. 

When deploying the Gauss-law ansatz to quantitatively investigate the in-medium modification of the potential, we find that the lattice extracted values of $\Re[V]$ are well reproduced by an analytic fit that depends only on a single temperature dependent parameter $m_D$. The values of $m_D$ reflect the qualitative behavior of $\Re[V]$ in that they vanish below $T_C$, appear to jump to a finite value at $T_C$ and continue to rise above. From an attempt to fit the temperature dependence of $m_D/T$ with a HTL inspired fitting formula, we learned that the downward trend in $m_D/T$ as one approaches the transition from above is a clear sign of non-perturbative physics, deviating from the purely perturbative predictions of HTL. A determination of $m_D/T$ at high temperatures should be attempted in order to ascertain, whether it actually continues to rise and eventually becomes compatible to the values extracted from EQCD effective field theory.

In order to consistently connect to continuum computations, the lattice results need to be extrapolated to vanishing lattice spacing. While our initial intent was to do so by using the results from $\beta=7$ lattices at $N_s=32$, our reassessment of the obtained results hints at the presence of significant finite volume artifacts entering the computed potential values. In particular we interpret the fact that one observes on these lattices a nearly constant $m_D$ from shortly above $T_C$ to far into the confined phase as a sign that the infrared cutoff $\Lambda_{\rm IR}$ has overwhelmed the dynamics at that point. We will therefore need to repeat the potential extraction at a comparable physical volume as used for the coarser lattices considered, corresponding to $N_s = 80$ at least. Additional inconsistencies in the assignment of the critical temperature and the occurrence of maximum Monte-Carlo autocorrelation times hints at the necessity to actually carry out the scale setting for this parameter set anew, which is planned to be undertaken soon.

Interestingly the behavior of $m_D$ found at $\beta=6.1$ is quite similar to that found in a study in dynamical QCD with $N_f=2+1$ dynamical quark flavors based on the ASQTAD action. It is the heavy pion mass $m_\pi\approx300$MeV, which in that case partially prevents the thermal activation of light quark degrees of freedom, leading to a qualitatively close outcome. For lighter quark degrees of freedom we expect that in the transition region around the pseudocritical temperature the onset of screening will proceed more softly than observed here. In order to determine the Debye mass parameter for use in phenomenology applications to heavy-ion collisions it is therefore paramount to proceed with the extraction of the potential also on current generation ensembles, where the pion masses take on almost physical values. 

\section*{Acknowledgments}

We thank N. Brambilla and P. Petreczky for insightful discussion. Calculations were performed on the in-house cluster at the ITP in Heidelberg and the SuperB cluster at EPFL. In addition we utilized the resources of the BWUniCluster at the KIT for generating the new ensembles at $\beta=7$. YB is supported by SNF grant PZ00P2-142524. This work is part of and supported by the DFG Collaborative Research Centre "SFB 1225 (ISOQUANT)"

\end{document}